\documentclass[fleqn,usenatbib]{mnras}

\usepackage[T1]{fontenc}

\DeclareRobustCommand{\VAN}[3]{#2}
\let\VANthebibliography\thebibliography
\def\thebibliography{\DeclareRobustCommand{\VAN}[3]{##3}\VANthebibliography}

\usepackage{graphicx}	
\usepackage{amsmath}	
\usepackage{amssymb}	

\usepackage{siunitx}
\usepackage{filecontents}
\usepackage{subfigure}
\usepackage{tikz}
\usepackage{multirow}
\usepackage{xcolor}
\usepackage{multicol}

\newcommand{\Msun}{\mbox{$M_{\odot}$}}
\newcommand{\logHHe}{\mbox{$\log\textnormal{(H/He)}$}}
\newcommand{\Teff}{\mbox{$T_\mathrm{eff}$}}
\newcommand{\kms}{\mbox{km\,s$^{-1}$}}

\title[Ultra-cool DZ white dwarfs]{Spectral analysis of ultra-cool white dwarfs polluted by planetary debris}

\author[A. K. Elms]{Abbigail K. Elms,$^{1}$\thanks{Contact e-mail: \href{mailto:Abbigail.Elms@warwick.ac.uk}{Abbigail.Elms@warwick.ac.uk}}
{Pier-Emmanuel Tremblay,$^{1}$}
{Boris T. G{\"a}nsicke,$^{1}$}
{Detlev Koester,$^{2}$}
\newauthor
{Mark A. Hollands,$^{3}$}
{Nicola Pietro Gentile Fusillo,$^{4}$}
{Tim Cunningham$^{1}$ and}
{Kevin Apps$^{5}$}
\\
$^{1}$Department of Physics, University of Warwick, Coventry, CV4 7AL, UK\\
$^{2}$Institut f{\"u}r Theoretische Physik und Astrophysik, University of Kiel, 24098 Kiel, Germany\\
$^{3}$Department of Physics and Astronomy, University of Sheffield, Sheffield, S3 7RH, UK\\
$^{4}$European Southern Observatory, Karl-Schwarzschild-Str 2, D-85748 Garching, Germany\\
$^{5}$Independent scholar, UK
}

\date{Accepted XXX. Received YYY; in original form ZZZ}

\pubyear{2022}

\usepackage{newtxtext,newtxmath}

\begin{document}
\label{firstpage}
\pagerange{\pageref{firstpage}--\pageref{lastpage}}
\maketitle

\begin{abstract}
We identify two ultra-cool ($\Teff < 4000$\,K) metal-polluted (DZ) white dwarfs WD\,J2147$-$4035 and WD\,J1922$+$0233 as the coolest and second coolest DZ stars known to date with $\Teff \approx 3050$\,K and $\Teff \approx 3340$\,K, respectively. Strong atmospheric collision-induced absorption (CIA) causes the suppression of red optical and infra-red flux in WD\,J1922$+$0233, resulting in an unusually blue colour given its low temperature. WD\,J2147$-$4035 has moderate infra-red CIA yet has the reddest optical colours known for a DZ white dwarf. Microphysics improvements to the non-ideal effects and CIA opacities in our model atmosphere code yields reasonable solutions to observations of these ultra-cool stars. WD\,J2147$-$4035 has a cooling age of over 10\,Gyr which is the largest known for a DZ white dwarf, whereas WD\,J1922$+$0233 is slightly younger with a cooling age of 9\,Gyr. Galactic kinematics calculations from precise \textit{Gaia} EDR3 astrometry reveal these ultra-cool DZ stars as likely members of the Galactic disc thus they could be pivotal objects in future studies constraining an upper age limit for the disc of the Milky Way. We present intermediate-resolution spectroscopy for both objects, which provides the first spectroscopic observations of WD\,J2147$-$4035. Detections of sodium and potassium are made in both white dwarfs, in addition to calcium in WD\,J1922$+$0233 and lithium in WD\,J2147$-$4035. We identify the magnetic nature of WD\,J2147$-$4035 from Zeeman splitting in the lithium line and also make a tentative detection of carbon, so we classify this star as DZQH. WD\,J1922$+$0233 likely accreted planetary crust debris, while the debris composition that polluted WD\,J2147$-$4035 remains unconstrained.
\end{abstract}

\begin{keywords}
white dwarfs -- stars: atmospheres -- stars: abundances -- methods: data analysis -- methods: analytical
\end{keywords}



\section{Introduction}
Over 97\,per\,cent of main-sequence stars ($M/\Msun$ $\lesssim$ 8) in the Milky Way galaxy will end their stellar evolution as white dwarfs, making these the most common type of stellar remnant. Typical white dwarfs consist of a dense carbon-oxygen (C/O) degenerate core of comparable radius to the Earth, surrounded by a thin envelope of residual hydrogen and helium left over from the progenitor’s post-main-sequence evolution. The degenerate core creates an efficient environment for electron conduction which makes it almost isothermal. 

Throughout their lives, white dwarfs slowly cool by radiating their residual internal thermal energy away through their thin non-degenerate envelopes \citep{Mestel1952, Althaus2010}. Thus the coolest white dwarfs, with ultra-cool effective temperatures (\Teff) $< 4000$\,K, tend to have the largest cooling ages depending on their atmospheric composition and radius (mass). Studying white dwarfs within the Milky Way allows us to calibrate stellar ages \citep{Fouesneau2019, RebassaMansergas2021} and probe the formation and evolution of the Galactic neighbourhood \citep{Winget1987, Rowell2013, Tremblay2014_SFH, Kilic2017, Fantin2019, Fantin2020}, with the analysis of cool remnants being particularly important as these provide constraints on its oldest stars \citep{Bergeron1997, Gianninas2015, Lam2020}.

Spectroscopy reveals the atmospheric composition of white dwarfs and allows us to separate them into spectral types. White dwarfs should only have hydrogen or helium present in their atmospheres due to high surface gravities ($\log g \approx 8$) and fast diffusion timescales causing heavy elements to sink below the photosphere \citep{Paquette1986, Koester2009}, yet metals are found to pollute 25~–~50\,per\,cent of all white dwarfs \citep{Zuckerman2003, Zuckerman2010, Koester2014}. It is now well established that these metals contaminate white dwarfs through the accretion of tidally disrupted rocky bodies which survived the final evolutionary stages of its host star \citep{Graham1990, Jura2003, Farihi2010, Gansicke2012, Veras2014, Cunningham2022}.

White dwarfs with $\Teff \lesssim 5000$\,K generally have featureless spectra as they are too cool to display hydrogen or helium absorption lines (DC spectral type), but can show strong metal lines (DZ spectral type) if polluted by planetary debris \citep{Sion1983}. 
The metal lines in these cool DZ white dwarfs can provide invaluable information on the physical conditions present in these dense, cool atmospheres in a way that is not possible with their featureless DC counterparts \citep{Blouin2019}. However, the extreme high density ($\gtrsim 0.1\,\mathrm{g}\,\mathrm{cm}^{-3}$) in the atmosphere of these remnants affects their observed photometry and spectroscopy which complicates assessments of their \Teff, mass and age. 

The atmospheric collision induced absorption (CIA) of H$_{2}$-H$_{2}$, H$_{2}$-He, H-He, H$_{2}$-H and He-He-He in cool white dwarfs can greatly alter their spectral energy distributions (SEDs), leading them to display optical and near-infrared (NIR) colours that can be much bluer than a blackbody \citep{Bergeron1994, Hansen1998, Blouin2017}. CIA opacities and the treatment of charged and neutral particle interactions must therefore be carefully considered within cool white dwarf model atmosphere codes. As the physics of these extreme conditions is not yet fully understood, uncertainties remain on mass and age estimates \citep{Bergeron_Leggett2002, Gianninas2015, Kilic2020, Hollands2021, Kaiser2021, Bergeron2022}. Regardless of \Teff, white dwarfs with strong CIA opacity are designated as infrared-faint \citep[IR-faint;][]{Kilic2020}.

The spacecraft {\it Gaia} \citep{Gaia2016} measures the precise astrometric and photometric quantities of stars in the Milky Way and has allowed the recent identification of more than 300\,000 new white dwarf candidates \citep{GF2019,GF2021}. Spectroscopic follow-ups \citep{Tremblay2020,Kaiser2021} have revealed several new cool DZ stars, which led to the first detections of lithium and potassium in the atmosphere of white dwarfs and different propositions to explain the intriguingly high abundances of these elements compared to calcium and sodium, including the accretion of planetary crusts and lithium-enhanced primordial gas \citep{Hollands2021,Kaiser2021}. Distinguishing between these two scenarios now requires an enlarged sample of cool DZ white dwarfs as well as a more accurate characterisation of their ages.

This work focuses on spectroscopic observations and analysis of two ultra-cool DZ white dwarfs WD\,J214756.59$-$403527.79 (hereafter WD\,J2147$-$4035) and WD\,J192206.20$+$023313.29 (hereafter WD\,J1922$+$0233). These were first identified as white dwarf candidates from \textit{Gaia} \citep{GF2019, GF2021}. \citet{Apps2021} re-identified WD\,J2147$-$4035 as a nearby star with unusual colours and speculated that it is an extremely cool white dwarf. WD\,J1922$+$0233 was previously confirmed as a DZ in \citet{Tremblay2020} while WD\,J2147$-$4035 is spectroscopically identified as a DZQH in this work -- we broadly refer to this star as a member of the DZ class throughout this paper. 

In Section~\ref{sec:WDsample}, we present spectroscopic and photometric observations of WD\,J2147$-$4035 and WD\,J1922$+$0233. These are put into the context of a sample of \textit{Gaia} white dwarfs within 100\,pc, including a subsample of DZ white dwarfs with $\Teff < 5000$\,K, in Section~\ref{sec:Similarly cool DZ white dwarfs}. We describe the microphysics improvements to our model atmospheres which allow us to derive the \Teff, $\log g$, mass, cooling age and chemical abundances of WD\,J2147$-$4035 and WD\,J1922$+$0233 in Sections~\ref{sec:model-atmospheres}--\ref{sec:Spectroscopic_observations}. In Section~\ref{sec:Discussion} we discuss our results and conclude in Section~\ref{sec:Conclusions}.

\section{Observations}
\label{sec:WDsample}

\begin{table*}
    \centering
	\caption{Astrometry for the white dwarfs in our DZ subsample. WD\,J2147$-$4035 and WD\,J1922$+$0233 are the main focus of this work, while the remaining six stars are used for comparison. The \textit{Gaia} EDR3 parallax ($\varpi$) values listed have been corrected to account for the zero point offset following \citet{Lindegren2021}. Proper motions, $\mu$, are given in the right ascension ($\alpha$) and declination ($\delta$) directions. We calculated the tangential velocity, $v_{\perp}$, and Galactic velocity components of all eight stars in this work, where $U$ indicates motion radially away from the Galactic centre, $V$ is in the direction of the Galaxy's rotation and $W$ is perpendicular to the disc. We have assumed zero radial velocity in the calculations of $U$, $V$ and $W$. Values are given in the J2016.0 epoch.}
	\label{tab:astrometry}
	\resizebox{\textwidth}{!}{\begin{tabular}{lcccccccccc}
		\hline
		\hline
		& RA & Dec & $\varpi$ & Distance & $\mu_\alpha\,\textnormal{cos}(\delta)$ & $\mu_\delta$ & $v_{\perp}$ & $U$ & $V$ & $W$ \\
		&  &  & (mas) & (pc) & (mas\,yr$^{-1}$) & (mas\,yr$^{-1}$) & (\kms) & (\kms) & (\kms) & (\kms)\\
		\hline
		\hline
		WD\,J2147$-$4035 & 21:47:56.59 & $-$40:35:27.78 & 35.79 $\pm$ 0.49 & 27.94 $\pm$ 0.38 & $-$84.11 $\pm$ 0.42 & $-$112.39 $\pm$ 0.42 & 18.59 $\pm$ 0.26 & $-$9.80 & $-$13.65 & 7.96\\
		WD\,J1922$+$0233 & 19:22:06.20 & +02:33:13.29 & 25.36 $\pm$ 0.27 & 39.43 $\pm$ 0.41 & 69.66 $\pm$ 0.25 & $-$29.16 $\pm$ 0.24 & 14.11 $\pm$ 0.15 & 1.83 & 0.07 & $-$14.00\\
		\hline
		WD\,J1824$+$1213 & 18:24:58.44 & +12:13:06.82 & 25.00 $\pm$ 0.18 & 40.00 $\pm$ 0.29 & $-$280.11 $\pm$ 0.18 & $-$1078.43 $\pm$ 0.17 & 211.27 $\pm$ 1.54 & $-$141.95 & $-$151.35 & $-$39.74\\
		WD\,J1330$+$6435 & 13:30:01.16 & +64:35:23.71 & 11.52 $\pm$ 0.78 & 86.79 $\pm$ 5.51 & $-$110.63 $\pm$ 1.51 & $-$23.30 $\pm$ 1.38 & 46.51 $\pm$ 3.02 & 30.45 & $-$32.86 & 12.51\\
		WD\,J1644$-$0449 & 16:44:17.01 & $-$04:49:47.71 & 12.65 $\pm$ 0.87 & 79.07 $\pm$ 5.07 & 80.58 $\pm$ 0.95 & $-$45.40 $\pm$ 0.75 & 34.67 $\pm$ 2.25 & $-$13.98 & 5.44 & $-$31.25\\
		WD\,J2356$-$2054 & 23:56:44.76 & $-$20:54:53.77 & 15.28 $\pm$ 0.58 & 65.46 $\pm$ 2.50 & $-$295.49 $\pm$ 0.57 & $-$239.37 $\pm$ 0.42 & 118.01 $\pm$ 4.33 & $-$114.85 & $-$24.96 & 10.58\\
		WD\,J2317$+$1830 & 23:17:26.73 & +18:30:52.75 & 26.40 $\pm$ 0.31 & 37.89 $\pm$ 0.44 & $-$33.85 $\pm$ 0.32 & $-$452.81 $\pm$ 0.26 & 81.55 $\pm$ 0.94 & $-$45.36 & $-$40.84 & $-$54.08\\
		WD\,J1214$-$0234 & 12:14:56.38 & $-$02:34:02.83 & 26.26 $\pm$ 0.12 & 38.09 $\pm$ 0.17 & 358.65 $\pm$ 0.14 & $-$419.28 $\pm$ 0.09 & 99.61 $\pm$ 0.44 & $-$92.41 & $-$24.09 & $-$28.33\\
		\hline
	\end{tabular}}
\end{table*}

\begin{figure*}
    \centering
    \includegraphics[width=2\columnwidth]{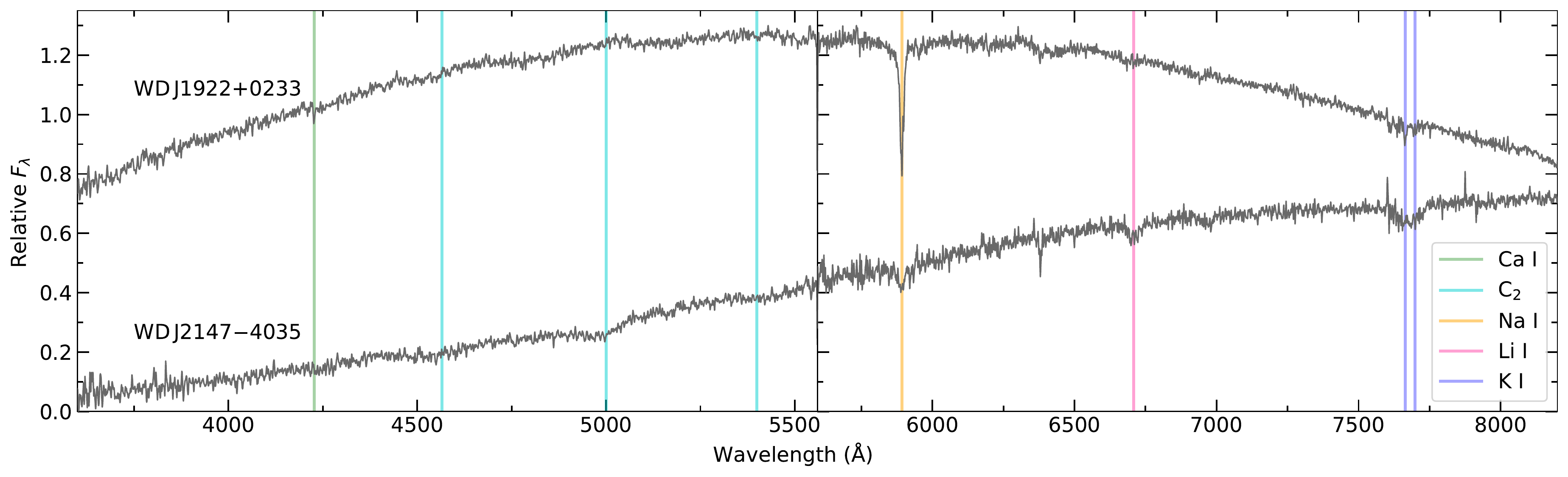}
\caption{Optical spectra (grey) from the VLT X-Shooter spectrograph for WD\,J2147$-$4035 and WD\,J1922$+$0233, comprised of observations taken with the UVB (left panel) and VIS (right panel) arms. Spectral lines of \ion{Ca}{i} (4227\,\AA), \ion{Na}{i} D (5893\,\AA), \ion{Li}{i} (6708\,\AA) and \ion{K}{i} (7665\,\AA\, and 7699\,\AA) are highlighted with coloured vertical bars to show metal detections of \ion{Na}{i} D and \ion{K}{i} in both white dwarfs, \ion{Ca}{i} in WD\,J1922$+$0233 and \ion{Li}{i} in WD\,J2147$-$4035. We made a tentative detection of carbon in WD\,J2147$-$4035 as the three strongest C$_2$ Swan band systems (aqua vertical lines) blueshifted to the centroid wavelengths measured in the DQpecP star LP\,351$-$42 are in excellent agreement to the observed broad, rounded absorption features. $F_\lambda$ is the flux per unit wavelength. Spectra are convolved by a Gaussian with a FWHM of 2\,\AA\, for clarity. Telluric absorption from the Earth’s atmosphere are evident in the region $\approx$ 7600 – 7640\,\AA\, in both spectra.}
\label{fig:VLT_spectra}
\end{figure*}

The \textit{Gaia} Early Data Release 3 \citep[EDR3;][]{GaiaCollab2021_summary} astrometry of WD\,J2147$-$4035 and WD\,J1922$+$0233 is given in Table~\ref{tab:astrometry}. We have corrected the parallax values for all objects to account for the zero point offset following \citet{Lindegren2021}.

\citet{Tremblay2020} conducted the first spectroscopic observations of WD\,J1922$+$0233 with the Gran Telescopio Canarias (GTC) Optical System for Imaging and low-Intermediate-Resolution Integrated Spectroscopy \citep[OSIRIS;][]{Cepa2000, Cepa2003} spectrograph. \citet{Tremblay2020} speculated that WD\,J1922$+$0233 is the first detected ultra-cool DZ white dwarf that exhibits strong optical CIA\footnote{\citet{Blouin2018b} have previously identified a cool DZ white dwarf SDSS\,J0804+2239 with strong near-IR CIA.}. Given the limited wavelength coverage of the observations, the sodium D-line was the only line detected. The first spectroscopic observations of WD\,J2147$-$4035 are presented in this work. It was selected as a high-confidence white dwarf candidate from \citet{GF2021} as part of a spectroscopic survey of the southern 40\,pc sample (O'Brien et al., in preparation).

We observed WD\,J2147$-$4035 and WD\,J1922$+$0233 on 2021 June 5-7 with X-shooter \citep{Vernet2011} on the Very Large Telescope (VLT) of the European Southern Observatory. We used slit widths of $1.0$, $0.9$, and $0.9$\,arcsec in the blue (UVB, $3000$~--~$5600$\,\AA, $R=5400$), visual (VIS, $5500$~--$~10\,200$\,\AA, $R=8900$) and NIR ($10\,200$~--~$24\,800$\,\AA, $R=5600$) arms, respectively. The exposure times in the UVB, VIS, and NIR arms were $2 \times 1800$, $2 \times 1750$, and $12 \times 300$\,s. The data were reduced following standard practices and using the \texttt{Reflex} pipeline \citep{Freudling2013}. The flux calibration was carried out using observations of the pure-hydrogen white dwarf LTT\,7987, obtained with the identical instrument setup as the science spectroscopy, and telluric correction was performed using \texttt{molecfit} \citep{Smette2015, Kausch2015}. The X-Shooter spectra for WD\,J2147$-$4035 and WD\,J1922$+$0233 are shown in Figure~\ref{fig:VLT_spectra}, where we have highlighted spectral lines of metals detected in one or both spectra (see Section~\ref{sec:Spectroscopic_observations}) with coloured vertical bars: calcium, sodium and potassium are detected in WD\,J1922$+$0233; sodium, lithium, potassium, and tentatively carbon, are detected in WD\,J2147$-$4035. The spectra obtained from the NIR arms of X-shooter for both stars are not used in this paper as they had insufficient flux for any scientific application.
 
Tables~\ref{tab:WDJ2147photometry} and~\ref{tab:WDJ1922photometry} display the available survey photometric data of WD\,J2147$-$4035 and WD\,J1922$+$0233, respectively. Corrections to the \textit{Gaia} EDR3 $G$ magnitude measurements were applied following the procedure from \citet{Lindegren2021}, with the corrected values shown in the tables. WD\,J1922$+$0233 has Panoramic Survey Telescope and Rapid Response System \citep[Pan-STARRS;][]{Chambers2016, Flewelling2020} DR2 photometry whereas WD\,J2147$-$4035 does not; instead, we used $grizY$ photometry from the Dark Energy Camera \citep[DECam;][]{DES2015_2} from DR1 of the Dark Energy Survey \citep[DES;][]{DES2016_1, Abbott2018} located at the Cerro Tololo Inter-American Observatory (CTIO), as this has similar bandpasses to Pan-STARRS. WD\,J2147$-$4035 also has NIR photometry from DR6 of the Visible and Infrared Survey Telescope for Astronomy (VISTA) Hemisphere Survey \citep[VHS;][]{McMahon2013} and the Wide-field Infrared Survey Explorer \citep[WISE;][]{WISE2010} CatWISE2020 catalogue \citep{Marocco2021}.

\begin{table}
	\centering
	\caption{Optical and IR photometry for WD\,J2147$-$4035. The \textit{Gaia} $G$ magnitude has been corrected following the procedure from \citet{Lindegren2021}.}
	\label{tab:WDJ2147photometry}
	\begin{tabular}{llc}
		\hline
		\hline
		Survey & Filter & Magnitude\\
		& &  (mag)\\
		\hline
		\hline
		\multirow{3}{1cm}{\textit{Gaia}} & $G$ & 19.959 $\pm$ 0.008 \\
		& $G_{\rm BP}$ & 20.949 $\pm$ 0.089 \\
		& $G_{\rm RP}$ & 19.024 $\pm$ 0.046 \\
		\hline
		\multirow{5}{1cm}{DES} & $g$ & 21.400 $\pm$ 0.020 \\
		& $r$ & 19.858 $\pm$ 0.007 \\
		& $i$ & 19.298 $\pm$ 0.007 \\
		& $z$ & 19.106 $\pm$ 0.010 \\
		& $Y$ & 18.901 $\pm$ 0.027 \\
		\hline
		\multirow{2}{1cm}{VHS} & $J$ & 17.581 $\pm$ 0.015 \\
		& $K$ & 17.625 $\pm$ 0.091 \\
		\hline
		\multirow{2}{1cm}{WISE} & $W1$ & 16.965 $\pm$ 0.050 \\
		& $W2$ & 16.717 $\pm$ 0.095\\
		\hline
	\end{tabular}
\end{table}

\begin{table}
	\centering
	\caption{Optical photometry for WD\,J1922$+$0233. The \textit{Gaia} $G$ magnitude has been corrected following the procedure from \citet{Lindegren2021}.}
	\label{tab:WDJ1922photometry}
	\begin{tabular}{llc}
		\hline
		\hline
		Survey & Filter & Magnitude\\
		& &  (mag)\\
		\hline
		\hline
		\multirow{3}{1.5cm}{\textit{Gaia}} & $G$ & 19.120 $\pm$ 0.004 \\
		& $G_{\rm BP}$ & 19.407 $\pm$ 0.036 \\
		& $G_{\rm RP}$ & 18.654 $\pm$ 0.032 \\
		\hline
		\multirow{5}{1.5cm}{\mbox{Pan-STARRS}} & $g$ & 19.588 $\pm$ 0.013 \\
		& $r$ & 19.056 $\pm$ 0.011 \\
		& $i$ & 18.937 $\pm$ 0.023 \\
		& $z$ & 19.099 $\pm$ 0.038 \\
		& $y$ & 19.467 $\pm$ 0.020 \\
		\hline
	\end{tabular}
\end{table}

\section{Sample of cool DZ white dwarfs}
\label{sec:Similarly cool DZ white dwarfs}
Several other metal-polluted white dwarfs with $\Teff < 5000$\,K are found in the literature, from which we selected a subsample to use as a comparison to WD\,J2147$-$4035 and WD\,J1922$+$0233 in terms of astrometric, photometric and atmospheric parameters. We limited our subsample to include spectroscopically confirmed DZ stars with: $\Teff < 5000$\,K and lithium and/or potassium detections; or, $\Teff < 4000$\,K within error bars. This selection criteria yielded WD\,J235645.10$-$205449.94 (hereafter WD\,J2356$-$2054\footnote{Also known as WD\,J2356$-$209 in \citet{Blouin2019}.}) from \citet{Blouin2019}, WD\,J164417.02$-$044947.71 (hereafter WD\,J1644$-$0449) from \citet{Kaiser2021} and four white dwarfs from \citet{Hollands2021}: WD\,J182458.45$+$121316.82, WD\,J133001.17$+$643523.69, WD\,J231726.74$+$183052.75 and WD\,J121456.38$-$023402.84 (hereafter WD\,J1824$+$1213, WD\,J1330$+$6435, WD\,J2317$+$1830 and WD\,J1214$-$0234, respectively). Together with WD\,J2147$-$4035 and WD\,J1922$+$0233, these eight white dwarfs comprise our subsample and their astrometry is given in Table~\ref{tab:astrometry}.

\subsection{\textit{Gaia} white dwarfs within 100 pc}
\label{sec:Gaia white dwarfs within 100 pc}
To determine the unique nature of WD\,J2147$-$4035 and WD\,J1922$+$0233, we employed Hertzsprung–Russell diagrams (HRDs) to compare them to the \textit{Gaia} EDR3 white dwarf sample within 100\,pc of the Sun and Sloan Digital Sky Survey \citep[SDSS;][]{York2000} footprint from \citet{GF2021}. We imposed selection criteria to only return objects which have SDSS $ugriz$ photometry and are within 100\,pc, which resulted in a sample of 5613 objects. We found 2865 (51.0\,per\,cent)\footnote{The total number of objects with a spectrum in the 100\,pc sample and SDSS footprint is 2917 (52.0\,per\,cent). The total number of confirmed white dwarfs (including main white dwarf spectral types, `Other SpT', all binaries and CVs) in the 100\,pc sample and SDSS footprint is 2887 (51.4\,per\,cent).} of these 5613 stars have published spectral classifications of white dwarf spectral types \citep{Sion1983}. All binary spectral types are excluded from the sample of 2865 white dwarfs. All spectral classifications were obtained from the \citet{GF2021} \textit{Gaia}$-$SDSS spectroscopic catalogue, the literature \citep{Kilic2006, Kilic2010, Kilic2020} or the Montreal White Dwarf Database \citep[MWDD;][]{Dufour2017}.

\begin{figure*}
	\subfigure[]{\includegraphics[width=\columnwidth]{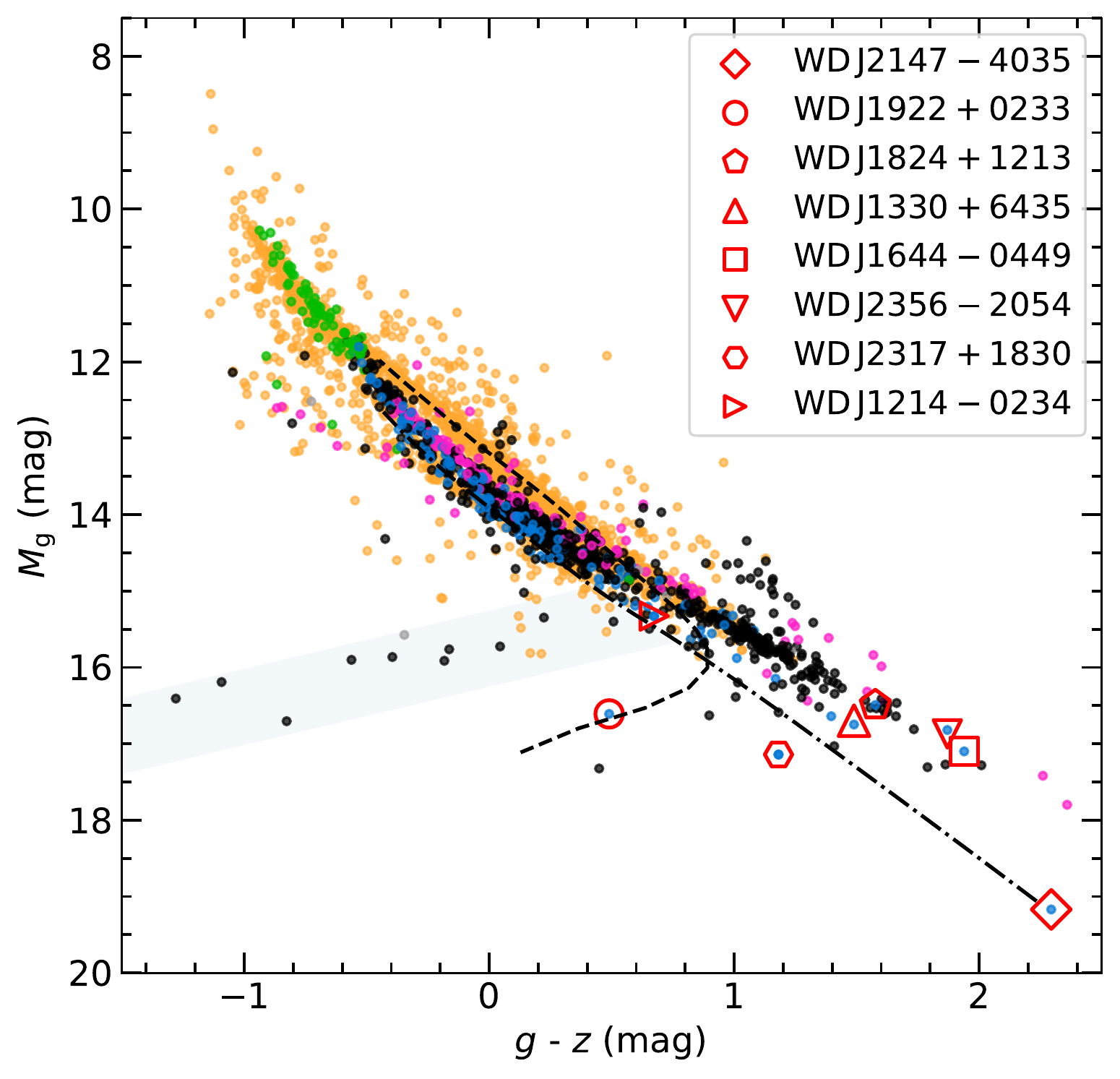}\label{fig:PS_gz}}
	\subfigure[]{\includegraphics[width=\columnwidth]{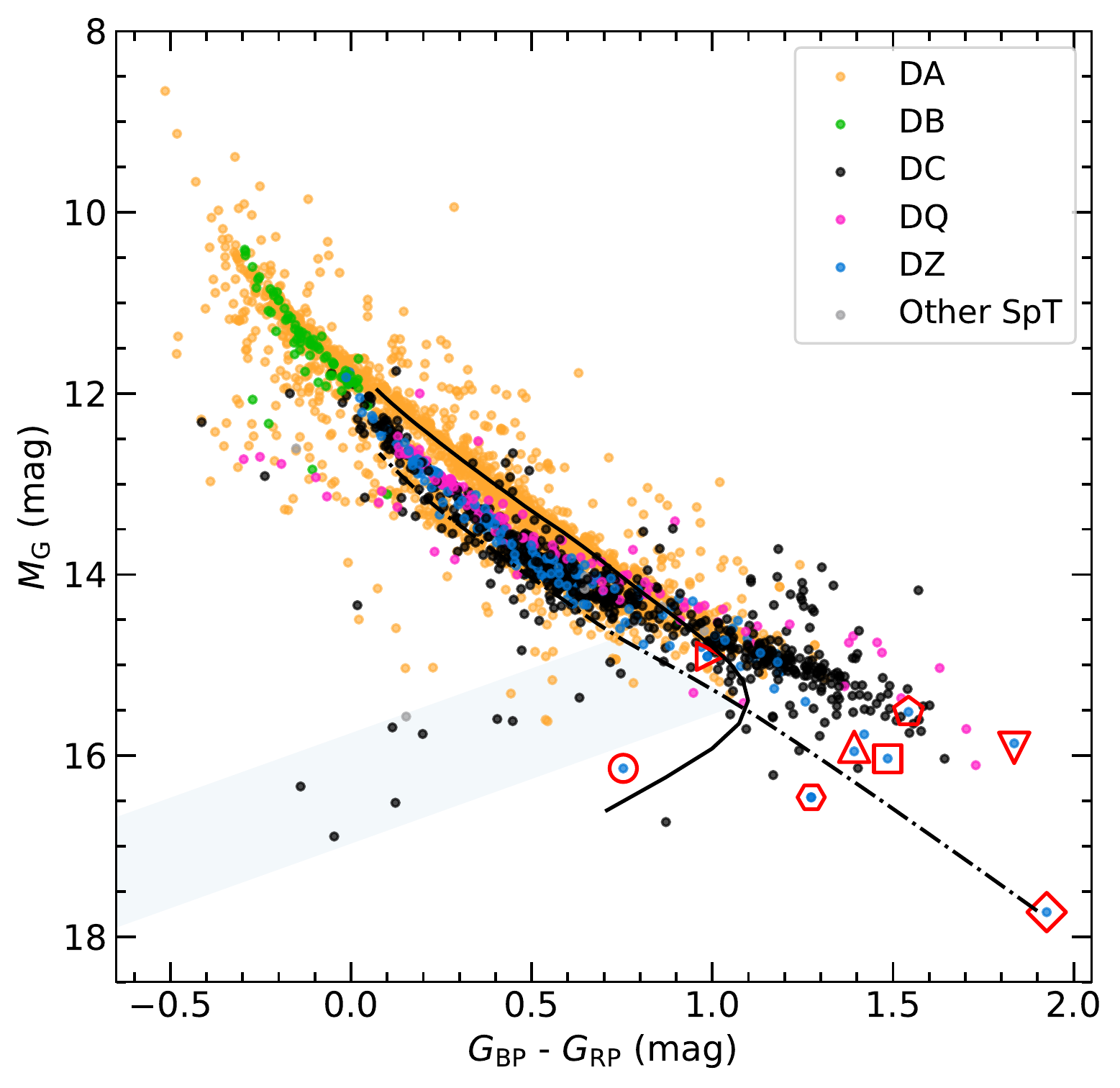}\label{fig:GaiaGR}}
    \caption{Hertzsprung-Russell diagrams displaying our spectroscopically confirmed 100\,pc white dwarf sample of 2871 stars, including the 2865 white dwarfs from the \citet{GF2021} \textit{Gaia} EDR3 100\,pc SDSS footprint and the DZ subsample stars defined in this work (see text), using photometry from (a) Pan-STARRS and (b) \textit{Gaia}. The DZ subsample is indicated by open red markers. DECam photometry is used for WD\,J2147$-$4035 in (a). The best-fitting indicative model cooling sequences, between $2750 < \Teff < 10\,000$\,K, for WD\,J2147$-$4035 is with $\log g = 8.25$ and $\logHHe = -5.0$ (dashed-dotted line) and for WD\,J1922$+$0233 is with $\log g=7.75$ and (a) $\logHHe = -2.5$ (dashed line) and (b) $\logHHe = -2.0$ (solid line). Objects on the ultra-blue sequence are indicated by the shaded blue region in both panels. The legends apply to both panels.}
    \label{fig:HR}
\end{figure*}

Narrow-band $grizy$ photometry from Pan-STARRS was subsequently obtained for the white dwarf sample within 100\,pc and the SDSS footprint, which complimented the broad bandpass photometry from \textit{Gaia} EDR3 in our analysis. The sample is displayed in the HRDs in Figure~\ref{fig:HR}, which is representative of the white dwarf cooling sequence. Absolute magnitudes are computed using \textit{Gaia} EDR3 parallaxes. Pan-STARRS photometry is used in Figure~\ref{fig:PS_gz} whereas \textit{Gaia} photometry is employed in Figure~\ref{fig:GaiaGR}. 

The DZ subsample stars were added to the main sample, except WD\,J1214$-$0234 and WD\,J2317$+$1830 as they are in the SDSS footprint, which increased our sample to 2871 white dwarfs within 100\,pc with spectral classifications; a breakdown of the number of white dwarfs with each spectral type is shown in Table~\ref{tab:sample_SpT}. The eight subsample DZ white dwarfs are plotted in Figure~\ref{fig:HR} and indicated with red open markers. WD\,J2147$-$4035 does not have Pan-STARRS photometry so we used DECam photometry instead.

\begin{table}
	\centering
	\caption{Breakdown of the identified main spectral types of the 2871 white dwarfs in the 100\,pc sample. This sample contains 2865 spectroscopically confirmed white dwarfs within 100\,pc and the SDSS footprint from \citet{GF2021}, including two DZ subsample stars defined in this work (see text) WD\,J2317$+$1830 and WD\,J1214$-$0234, and the remaining six DZ subsample stars. All binary spectral types and CVs are excluded. `Other SpT' are stars that are identified as white dwarfs but cannot be subclassified due to low-quality spectra, magnetic fields or peculiarities; additional observations or modelling is needed to clarify these spectral types.}
	\label{tab:sample_SpT}
	\begin{tabular}{lc}
		\hline
		\hline
		Spectral Type & Total Number \\
		\hline
		\hline
		DA & 1861 \\
		DB & 74 \\
		DC & 667 \\
		DQ & 117 \\
		DZ & 138 \\
		Other SpT & 8 \\
		\hline
		Subsample DZ & 6 \\
		\hline
	\end{tabular}
\end{table}

WD\,J2147$-$4035 and WD\,J1922$+$0233 occupy relatively isolated positions on the HRDs, suggesting they have unusual parameters and extreme natures (see Section~\ref{sec:nature} for further discussion). WD\,J1922$+$0233 has an unusually blue colour relative to its dim absolute magnitude and therefore could have strong CIA opacity, as suggested by the unusual spectral shape and red flux deficit in Figure~\ref{fig:VLT_spectra} \citep[see also][]{Tremblay2020}. WD\,J2147$-$4035 is an obvious outlier to the main white dwarf cooling sequence as it is the dimmest and reddest object out of the sample, which suggests it is an old remnant. Comparison with the local 40\,pc \textit{Gaia} sample reveals WD\,J2147$-$4035 as the intrinsically faintest spectroscopically confirmed white dwarf in the optical (O'Brien et al., in preparation). Also, the extreme red colour of WD\,J2147$-$4035 indicates it has more moderate CIA.

WD\,J1214$-$0234 lies on the main white dwarf cooling sequence. WD\,J1644$-$0449, WD\,J1824$+$1213, WD\,J1330$+$6435 and WD\,J2317$+$1830 are dimmer and redder than the majority of objects in the sample, thus reside in the tail of the main white dwarf cooling sequence. The cool DZ WD\,J2356$-$2054 is an outlier in Figure~\ref{fig:GaiaGR} as it has an extremely red colour of $G_{\rm BP}~-~G_{\rm RP} = 1.84$\,mag; this quantity is only slightly smaller compared to WD\,J2147$-$4035. A larger photometric difference is measured using Pan-STARRS between the two stars, which have measurements of $g-z = 1.88$\,mag for WD\,J2356$-$2054 and $g-z = 2.29$\,mag for WD\,J2147$-$4035. WD\,J2356$-$2054 has an absolute magnitude with Pan-STARRS and \textit{Gaia} photometry comparable to the DZ stars in our subsample which cluster at the end of the main white dwarf cooling sequence, and is notably brighter than WD\,J2147$-$4035 in $M_{\rm g}$ by $\approx 2.3$\,mag and $M_{\rm G}$ by $\approx 1.9$\,mag. The combination of its absolute magnitude and slightly bluer $g-z$ colour compared to its $G_{\rm BP}~-~G_{\rm RP}$ colour makes WD\,J2356$-$2054 occupy a less unusual position in Figure~\ref{fig:PS_gz} than Figure~\ref{fig:GaiaGR}.

Given the narrow Pan-STARRS bandpasses, metal lines in the DZ subsample and WD\,J2356$-$2054 could impact their positions in the HRD. However, we note that most warmer DZ white dwarfs sit on the main white dwarf cooling track. This suggests that it is the cool nature of these DZ white dwarfs that make them outliers in the HRD, rather than metal lines. This is confirmed by looking at the HRD of Figure\,\ref{fig:GaiaGR} using the much broader \textit{Gaia} filters.

\subsubsection{Ultra-blue sequence}
\label{sec:ultra-blue sequence}
A clear ultra-blue sequence consisting of mainly DC stars is seen in Figure~\ref{fig:HR} and indicated by the shaded blue region in both panels. This sequence was first extensively studied in \citet{Kilic2020}. White dwarfs on the ultra-blue sequence are designated as IR-faint as they have optical and NIR flux deficits and thus present strong CIA opacity, with models suggesting mixed H/He atmospheric compositions (see Section~\ref{sec:IR-faint}). 

The white dwarfs populating the ultra-blue sequence in Figure~\ref{fig:HR} have published \logHHe~$\sim-3.0$ and are modelled to be as cool as $\Teff \approx 2700$\,K \citep{Kilic2020}, although a more recent analysis by \citet{Bergeron2022} found a warmer lower limit of $\Teff \approx 3900$\,K. WD\,J003908.38$+$303539.76 is the bluest DC white dwarf on the ultra-blue sequence with $\logHHe = -3.4$ and $\Teff = 4605 \pm 77$\,K \citep{Bergeron2022}.

No white dwarf from our DZ subsample reside on the ultra-blue sequence. However, WD\,J1922$+$0233 resides parallel to the ultra-blue sequence so it could have turned off the main white dwarf cooling sequence later than the other objects in this population, possibly due to it having a slightly different \logHHe\ atmospheric abundance.

\section{Model atmospheres}
\label{sec:model-atmospheres}
The model atmosphere code of \citet{Koester2010}, with microphysics improvements explained in this section, was used to determine the parameters of WD\,J2147$-$4035 and WD\,J1922$+$0233. 

The atmospheric density of cool white dwarfs can become so high ($\gtrsim 0.1\,\mathrm{g}\,\mathrm{cm}^{-3}$) that it must be considered as a dense fluid rather than an ideal gas \citep{Kowalski_Saumon2004, Kowalski2006b, Kowalski2014, Blouin2017, Blouin2018a}. Deviations from the simple ideal equation-of-state (EOS) become important at these temperatures, especially for helium-rich models. Since the absorption coefficient of pure neutral helium is very small, these atmosphere models reach very high densities. We therefore utilised a non-ideal EOS taking into account the interaction of charged and neutral particles. For the description of the interaction between charged particles we followed \citet{Chabrier1998, Potekhin2000}, but used only the ion-ion interaction as described by the one-component-Coulomb plasma. The model of neutral interactions is based on the classical excluded volume model with hard spheres \citep[see e.g.][]{Hummer1988}, but modified arbitrarily to obtain the desired result of total ionisation in the density regime of about 3-8 g cm$^{-3}$ \citep{Saumon1995}. Together the charged and neutral particle interactions lead to a lowering of the energy for the ionisation of individual atoms and the dissociation of molecules. The non-ideal effects also have a contribution to the pressure and to other thermodynamic quantities like the adiabatic gradient.

We employed CIA opacities for H$_{2}$-H$_{2}$ \citep{Abel2011, Borysow2001}, H$_{2}$-He (\citealt{Abel2012} with scaling factor from \citealt{Blouin2017}), H-He \citep{Gustafsson2001} and H$_{2}$-H \citep{Gustafsson2003}. In principle there should also be a He-He-He CIA contribution \citep{Kowalski2014}. We implemented the numerical fit given by \citet{Kowalski2014} and find such a strong IR absorption in WD\,J2147$-$4035 that it is impossible to fit the photometry. A likely reason for this is that the maximum density \citet{Kowalski2014} used in the derivation of the numerical fit was $0.1\,\mathrm{g}\,\mathrm{cm}^{-3}$, and in the models shown the He-He-He CIA produces only a fairly minor change. However, in the photospheres of our models the densities reach $3\,\mathrm{g}\,\mathrm{cm}^{-3}$ ($n_{\mathrm{He}} = 4.5 \times 10^{23}$\,cm$^{-3}$), which is approximately the photospheric density in WD\,J2147$-$4035. As the absorption is proportional to the cube of the density, the extrapolation of the He-He-He opacity to the very high densities in our models is very likely not reliable. We have therefore switched off the effect from He-He-He CIA absorption in our models. For WD\,J1922$+$0233, the He-He-He contribution is negligible because of the much stronger contributions involving hydrogen.

\section{Stellar parameters}
\label{sec:stellar_params}

\begin{table*}
	\centering
	\caption{Atmospheric parameters for the white dwarfs in our DZ subsample. Parameters for WD\,J2147$-$4035 and WD\,J1922$+$0233 are derived in this work. The quoted uncertainties are purely of statistical nature as derived from the fits, and are much smaller than the model-dependent systematic uncertainties. The parameters for WD\,J1644$-$0449 are taken from \citet{Kaiser2021}, except for radius ($R$) and cooling age ($\tau$) which are calculated in this work. Similarly, the parameters for WD\,J2356$-$2054 are taken from \citet{Blouin2019}, except for radius ($R$) and white dwarf mass ($M_{\mathrm{WD}}$) which are calculated in this work. Parameters for WD\,J1824$+$1213, WD\,J1330$+$6435, WD\,J2317$+$1830 and WD\,J1214$-$0234 are from \citet{Hollands2021}. Observational upper limits of \logHHe\ are given where a tight constraint could not be derived.}
	\label{tab:atmospheric}
	\begin{tabular}{lcccccc} 
		\hline
		\hline
		& \Teff\ & $\log g$ & \logHHe & $R$ & $M_{\mathrm{WD}}$ & $\tau$ \\
		& (K) & (cm\,s$^{-2}$) & & ($\times$10$^{-5}$ R$_\odot$) & (M$_\odot$) & (Gyr)\\
		\hline
		\hline
		WD\,J2147$-$4035 & 3048 $\pm$ 35 & 8.195 $\pm$ 0.042 & < $-$5.66 & 1100 $\pm$ 32 & 0.69 $\pm$ 0.02 & 10.2 $\pm$ 0.2\\
		WD\,J1922$+$0233 & 3343 $\pm$ 54 & 8.000 $\pm$ 0.055 & $-$2.69 $\pm$ 0.17 & 1247 $\pm$ 26 & 0.57 $\pm$ 0.03 & 9.0 $\pm$ 0.2\\
		\hline
		WD\,J1824$+$1213 & 3350 $\pm$ 50 & 7.41 $\pm$ 0.07 & $-$0.07 $\pm$ 0.10 & 1741 $\pm$ 55 & 0.28 $\pm$ 0.03 & 5.5 $\pm$ 0.4\\
		WD\,J1330$+$6435$^a$ & 3660 $\pm$ 50 & 7.65 $\pm$ 0.14 & < $-$4.0 & 1524 $\pm$ 115 & 0.38 $\pm$ 0.07 & 6.0 $\pm$ 1.0\\
		WD\,J1644$-$0449 & 3830 $\pm$ 230 & 7.77 $\pm$ 0.23 & < $-$2.0 & 1404 $\pm$ 174 & 0.45 $\pm$ 0.12 & 6.8 $\pm$ 0.2\\
		WD\,J2356$-$2054 & 4040 $\pm$ 110 & 7.98 $\pm$ 0.07 & $-$1.5 $\pm$ 0.2 & 1263 $\pm$ 52 & 0.56 $\pm$ 0.04 & 8.0 $\pm$ 0.8\\
		WD\,J2317$+$1830 & 4210 $\pm$ 50 & 8.64 $\pm$ 0.03 & $-$0.09 $\pm$ 0.09 & 793 $\pm$ 21 & 1.00 $\pm$ 0.02 & 9.5 $\pm$ 0.2\\
		WD\,J1214$-$0234 & 4780 $\pm$ 50 & 7.97 $\pm$ 0.04 & $-$3.19 $\pm$ 0.18 & 1269 $\pm$ 29 & 0.55 $\pm$ 0.02 & 5.8 $\pm$ 0.2\\
		\hline
	\end{tabular}\\
\raggedright\footnotesize{$^a$Note that \citet{Blouin2019} found a higher \Teff\ of $4310 \pm 190$\,K and $\log g$ of $8.26 \pm 0.15$, which were used by \citet{Kaiser2021} to derive a white dwarf mass ($M_{\mathrm{WD}}$) of $0.74 \pm 0.10$\,M$_\odot$. These studies use different model atmospheres to \citet{Hollands2021}.}
\end{table*}

Representative cooling sequences which indicate approximate $\log g$ and \logHHe\ for WD\,J2147$-$4035 and WD\,J1922$+$0233 are shown in Figure~\ref{fig:HR} with the dashed, solid or dashed-dotted lines. The best-fitting cooling sequence for WD\,J2147$-$4035 is given in both HRDs by $\log g = 8.25$ and $\logHHe = -5.0$. The cooling sequences with $\log g = 7.75$ indicate the best-fit to WD\,J1922$+$0233 when using mixed atmospheres of $\logHHe = -2.5$ with Pan-STARRS photometry and $\logHHe = -2.0$ with \textit{Gaia} photometry. These cooling sequences serve as an indicative tool for deriving the atmospheric parameters of WD\,J2147$-$4035 and WD\,J1922$+$0233 using a model grid, as described in Section~\ref{sec:model-atmospheres}, and the full photometric data sets. 

Our model grid was used to predict synthetic photometry for both objects, which depends on the atmospheric parameters \Teff, $\log g$ and \logHHe\ abundance. We relied on the mass-radius relation of \citet{Bedard2020} to predict absolute magnitudes. We used an iterative procedure utilising the photometric (Tables~\ref{tab:WDJ2147photometry} and~\ref{tab:WDJ1922photometry}) and spectroscopic data sets to find the best-fitting atmospheric parameters and individual metal abundances of WD\,J2147$-$4035 and WD\,J1922$+$0233. We initially estimated atmospheric parameters using the photometric technique, where we fit the observed photometry to our grid of synthetic photometry which included metals over all bandpasses to minimise chi-square ($\chi^2$). Our minimisation technique utilised the non-linear least-squares Levenberg-Marquardt algorithm \citep{Press1986} and considered \Teff, $\log g$ and \logHHe\ as free parameters. We produced synthetic spectra at those fixed atmospheric parameters, then analysed the fit between the metal absorption lines in the synthetic and observed spectra. The metal abundances were adjusted to optimise the fit. Keeping these new metal abundances fixed, we produced a new grid of synthetic photometry and repeated the photometric fitting. This process was iterated until the best-fitting atmospheric parameters and metal abundances were found. The uncertainties of the best-fitting atmospheric parameters are dependent on one another and are computed from the covariance matrix of the model atmosphere fit, therefore they are statistical in nature and do not account for systematic uncertainty.

Our iterative procedure was most important for WD\,J2147$-$4035 as the metal abundances had a non-negligible effect on its atmospheric parameters. The reason for this effect is most likely due to the low atmospheric hydrogen content in the star, hence the metals provide additional free electrons and impact the photospheric structure. Alternatively, the inclusion of metals in the fit of WD\,J1922$+$0233 changed the synthetic photometry by less than 0.01\,mag, which is within $1\sigma$ and therefore not significant.

A lower limit of 0.05\,mag for photometric uncertainties was imposed in the fit of WD\,J2147$-$4035. This treatment prevented it from being dominated by the very small observed photometric uncertainties at optical wavelengths which may be underestimated considering the star's extremely dim magnitude, compared to the moderately large IR photometric uncertainties (see Table~\ref{tab:WDJ2147photometry}). The best-fitting model atmosphere for WD\,J2147$-$4035 included optimised fixed metal abundances (see Section~\ref{sec:Spectroscopic_observations}) with $\Teff = 3048\pm35$\,K, $\log g = 8.195\pm0.042$ and $\logHHe < -5.66$. Since a further decrease in the hydrogen content led only to minimal changes of the models, we would not exclude the possibility that our best-fitting \logHHe\ abundance is an upper limit instead of a real determination.

The \textit{Gaia} EDR3 $G$ magnitude uncertainties is an order of magnitude smaller than all other optical photometric uncertainties for WD\,J1922$+$0233 (see Table~\ref{tab:WDJ1922photometry}), therefore we added a common systematic uncertainty of 0.01\,mag to all measured photometric uncertainties for the fit. Despite the WD\,J1922$+$0233 \textit{Gaia} EDR3 parallax and uncertainty having a relative precision of 1.05\,per\,cent, we also included the parallax as a free parameter in the fit to propagate extra freedom into ${\Teff}$, $\log g$ and \logHHe, resulting in their uncertainties being more realistic. The best-fitting model atmosphere for WD\,J1922$+$0233 has $\Teff = 3343\pm54$\,K, $\log g = 8.000\pm0.055$ and $\logHHe = -2.69\pm0.17$. The best-fitting parallax is $25.330 \pm 0.265$, which has a percentage difference to the measured \textit{Gaia} parallax of only 0.126\,per\,cent.

After the completion of this work, a paper by \citet{Bergeron2022} found significant differences in the atmospheric parameters of WD\,J1922$+$0233, notably $\Teff = 4436\pm53$\,K, $\log g = 8.766$ and $\logHHe = -1.73$. However, the physics involved in our atmospheric models (see Section~\ref{sec:model-atmospheres}) and the ones used in \citet{Bergeron2022} differ slightly. Our models included the more recent H$_{2}$-He CIA calculations by \citet{Abel2012}, which should be superior to \citet{Jorgensen2000} used by \citet{Bergeron2022}. We also included non-ideal effects in the EOS. With our models, we found a satisfactory fit to the photometry of WD\,J1922$+$0233 for parameters significantly different from \citet{Bergeron2022}. A more detailed discussion is beyond the scope of this paper.

The best-fitting atmospheric parameters for our subsample are shown in Table~\ref{tab:atmospheric}. Note that the small uncertainties given for our parameters only include the statistical errors from our fits and no systematic errors from e.g. model uncertainties, so the real errors are likely larger.

\begin{figure}
    \centering
    \includegraphics[width=\columnwidth]{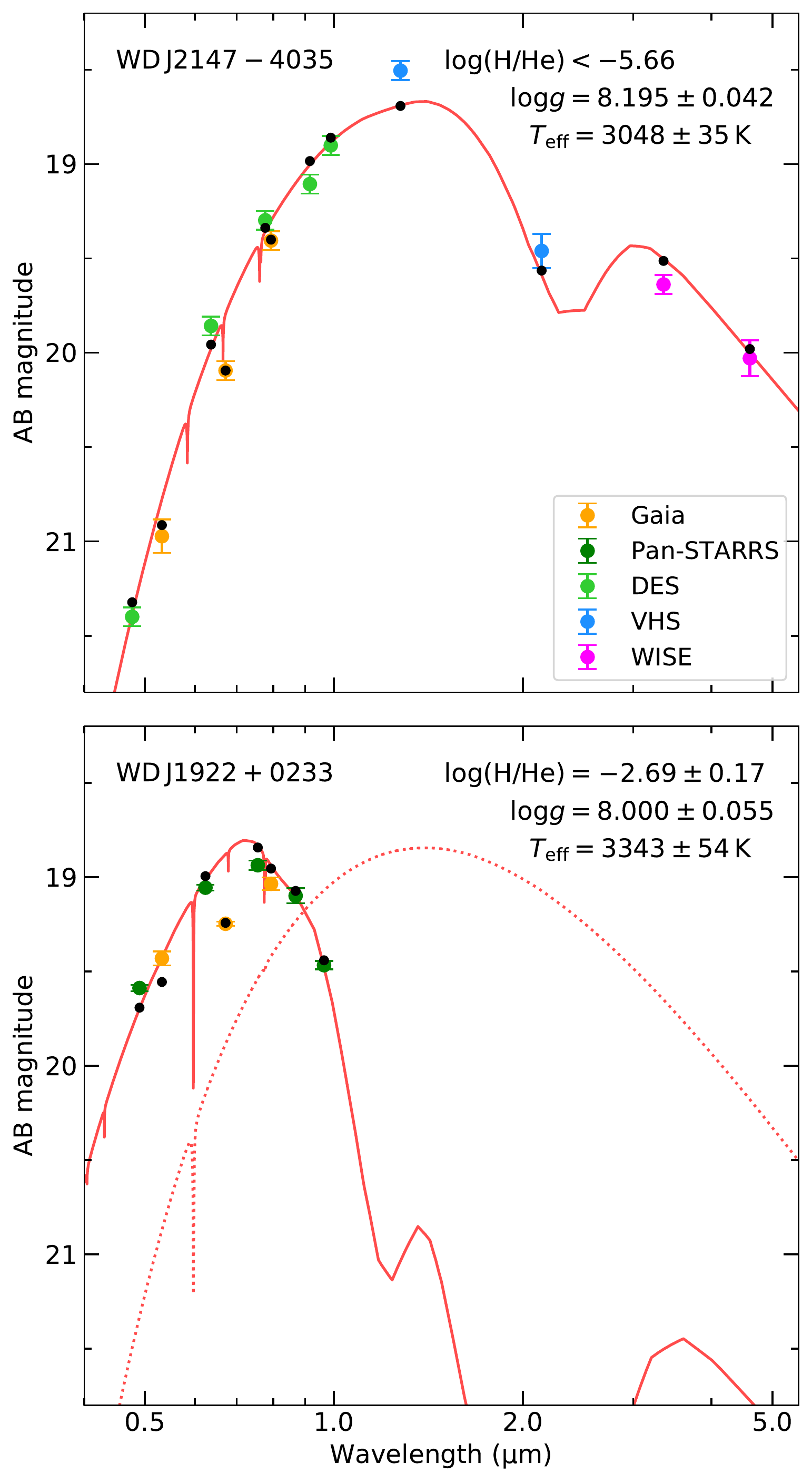}
\caption{Photometric fits between the observed (coloured circles with error bars) and synthetic (black circles) photometry for WD\,J2147$-$4035 (top panel) and WD\,J1922$+$0233 (bottom panel). Monochromatic model fluxes were computed using the best-fitting models including metals (see Section \ref{sec:Spectroscopic_observations}) and converted to AB magnitudes (red solid line) for each white dwarf. The predicted \textit{Gaia} EDR3 $G$, and to a lesser extent $G_{\rm BP}$ and $G_{\rm RP}$, magnitudes are visually offset from the predicted monochromatic magnitudes because of the very broad \textit{Gaia} bandpasses. All CIA opacities are removed in the best-fitting model with metals for WD\,J1922$+$0233 (red dotted line) to demonstrate the strong effect of CIA opacity on the photometry of this white dwarf. Error bars correspond to 1$\sigma$ uncertainties. The legend applies to both panels.}
\label{fig:photometric_fits}
\end{figure}

Monochromatic model fluxes were calculated with the best-fitting models including metals in the atmospheric structure calculations for WD\,J2147$-$4035 and WD\,J1922$+$0233. The best-fitting monochromatic model fluxes and synthetic photometry for both stars were converted to AB magnitudes and are shown in Figure~\ref{fig:photometric_fits} as a comparison to all available optical and NIR observed photometry. A reasonable fit is achieved between the observed and synthetic photometry for both objects. A visual offset is seen between the synthetic \textit{Gaia} EDR3 $G$, and to a lesser extent $G_{\rm BP}$ and $G_{\rm RP}$, magnitudes and our best-fitting models due to the very broad \textit{Gaia} bandpasses. We note the CIA opacities included in the model atmosphere code (see Section~\ref{sec:model-atmospheres}) have a strong effect on the quality of fit of the best-fitting model for WD\,J1922$+$0233 to observed photometry, where Figure~\ref{fig:photometric_fits} shows the best-fitting model (solid red line) compared to the identical model with all CIA opacities removed (dotted red line). The effect from CIA opacities in WD\,J1922$+$0233 is also clear from the blueward shifted peak of its SED to optical wavelengths, near $0.7\,\mu\mathrm{m}$ in this case, compared to when CIA opacities are removed. This shift represents the extreme deficit of IR flux due to CIA in this ultra-cool star, despite us lacking observational data in the IR. The SED peak of WD\,J2147$-$4035 is near $1.4\,\mu\mathrm{m}$ and is comparable to the SED peak of the best-fitting model of WD\,J1922$+$0233 when CIA opacities are removed, which suggests there are only mild effects from CIA opacity in this remnant.

The white dwarf mass ($M_{\mathrm{WD}}$) and cooling age ($\tau$) of WD\,J2147$-$4035 and WD\,J1922$+$0233 were derived using model evolutionary sequences with thin hydrogen layers (small total hydrogen masses of $1 \times 10^{-10}\,M_{\mathrm{H}}/M_{\mathrm{WD}}$, where $M_{\mathrm{H}}$ is the hydrogen mass), and C/O cores \citep{Bedard2020}, and are presented in Table~\ref{tab:atmospheric} with the rest of our subsample. WD\,J2147$-$4035 is relatively massive compared to other cool and ultra-cool remnants with $M_{\mathrm{WD}} = 0.69\pm0.02$\,\Msun\ and has the largest cooling age known for a DZ white dwarf of $\tau = 10.21\pm0.22$\,Gyr. We find $M_{\mathrm{WD}} =  0.57\pm0.03$\,\Msun\ for WD\,J1922$+$0233 and $\tau = 9.05\pm0.22$\,Gyr, making it a relatively old white dwarf in terms of cooling age. These white dwarfs have a very low temperature compared to the so-called truncation of the luminosity function at $\approx 4000$\,K, which is thought to correspond to $\tau \approx 10$\,Gyr and thus the oldest white dwarfs in the local Galactic disc \citep{Winget1987,Kilic2017}. White dwarfs with helium-rich atmospheres are thought to cool more rapidly than ones with hydrogen-rich atmospheres due to the early event of convective coupling \citep{Oppenheimer2001,Fontaine2001}, i.e. they will have a lower \Teff\ for the same cooling age. In addition, more massive white dwarfs, and therefore those with smaller radii according to the mass-radius relation of degenerate stars, cool faster than stars of more moderate mass, as they develop a crystallized core earlier so they reach the state of very small specific heat capacity values quicker thus have a rapid final cooling phase \citep{Fontaine2001}. The combination of these effects can help to explain the ultra-cool nature of both stars.

\section{Metal abundances}
\label{sec:Spectroscopic_observations}
The intermediate-resolution optical spectra from our VLT X-Shooter observations of WD\,J2147$-$4035 and WD\,J1922$+$0233 are shown in Figure~\ref{fig:VLT_spectra}. Strong lines of the \ion{Na}{i} D doublet (5893\,\AA) and \ion{K}{i} (7665\,\AA\, and 7699\,\AA) are seen in both spectra. We also detected \ion{Ca}{i} (4227\,\AA) in WD\,J1922$+$0233 and \ion{Li}{i} (6708\,\AA) in WD\,J2147$-$4035. We made a tentative detection of C$_2$ in WD\,J2147$-$4035 at $\approx 4570$\,\AA, 5000\,\AA\ and 5400\,\AA.

\begin{figure*}
    \centering
    \subfigure{\includegraphics[width=\columnwidth]{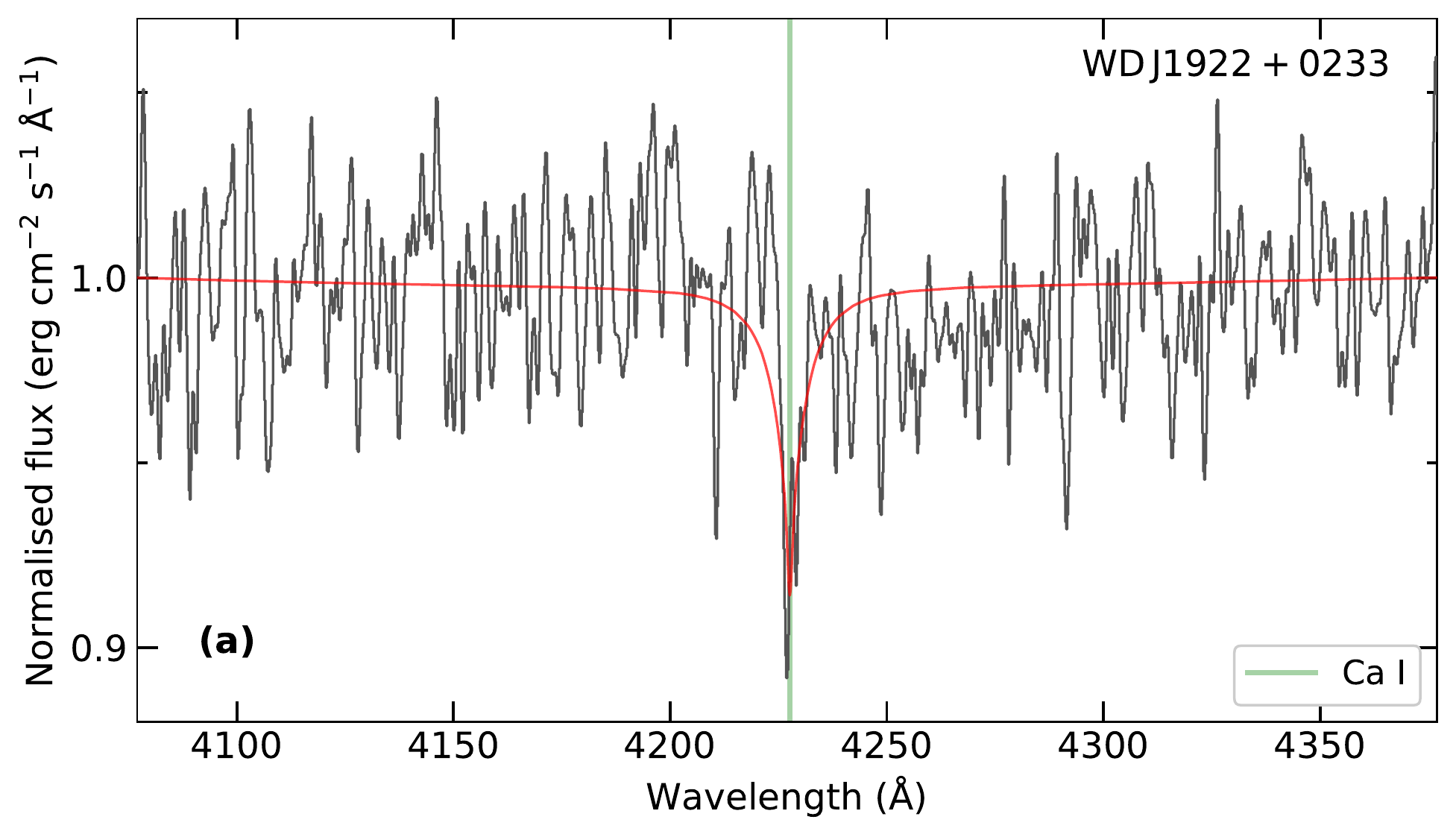}}
    \subfigure{\includegraphics[width=\columnwidth]{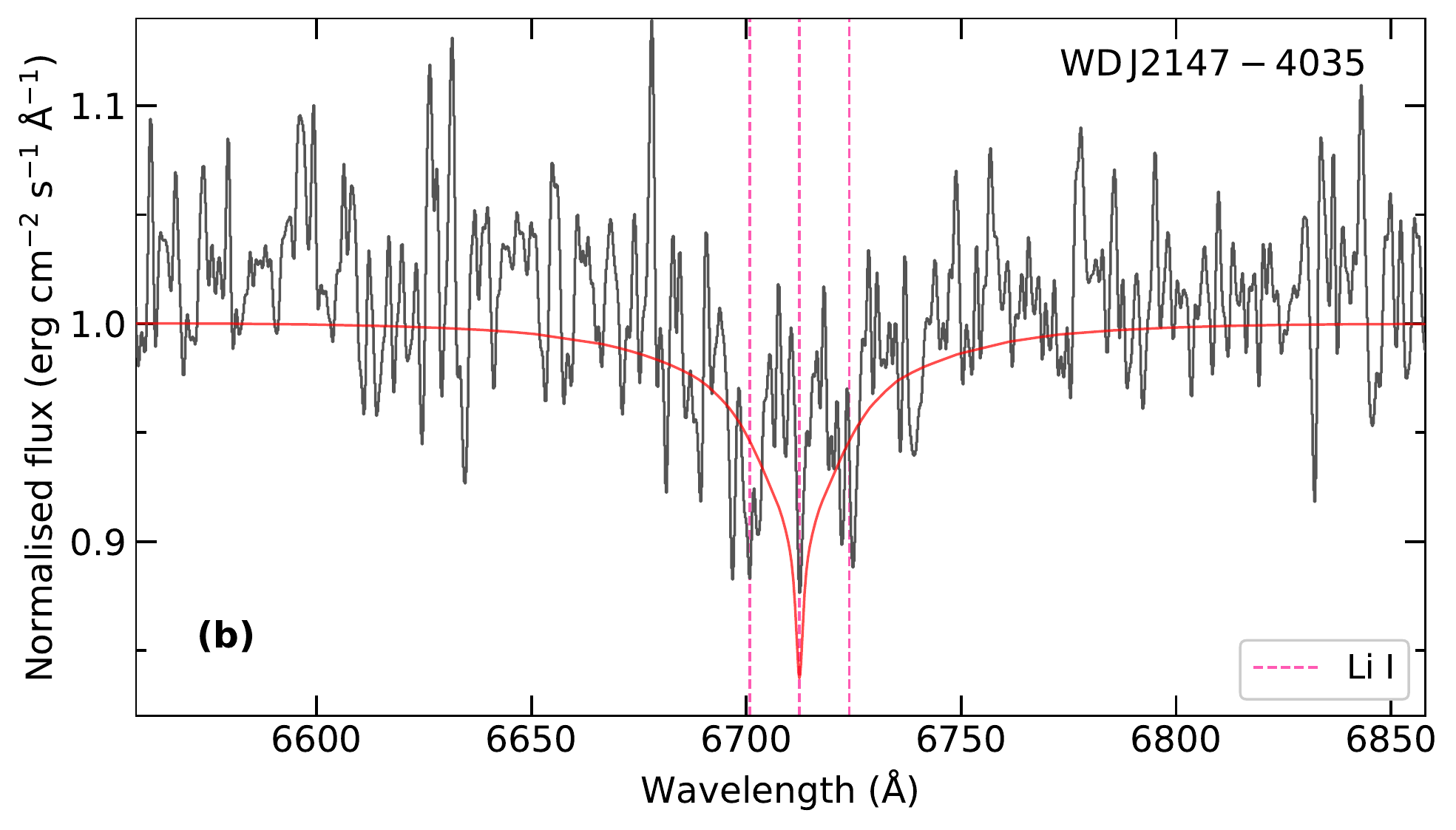}\label{fig:Li split line}}
    \subfigure{\includegraphics[width=\columnwidth]{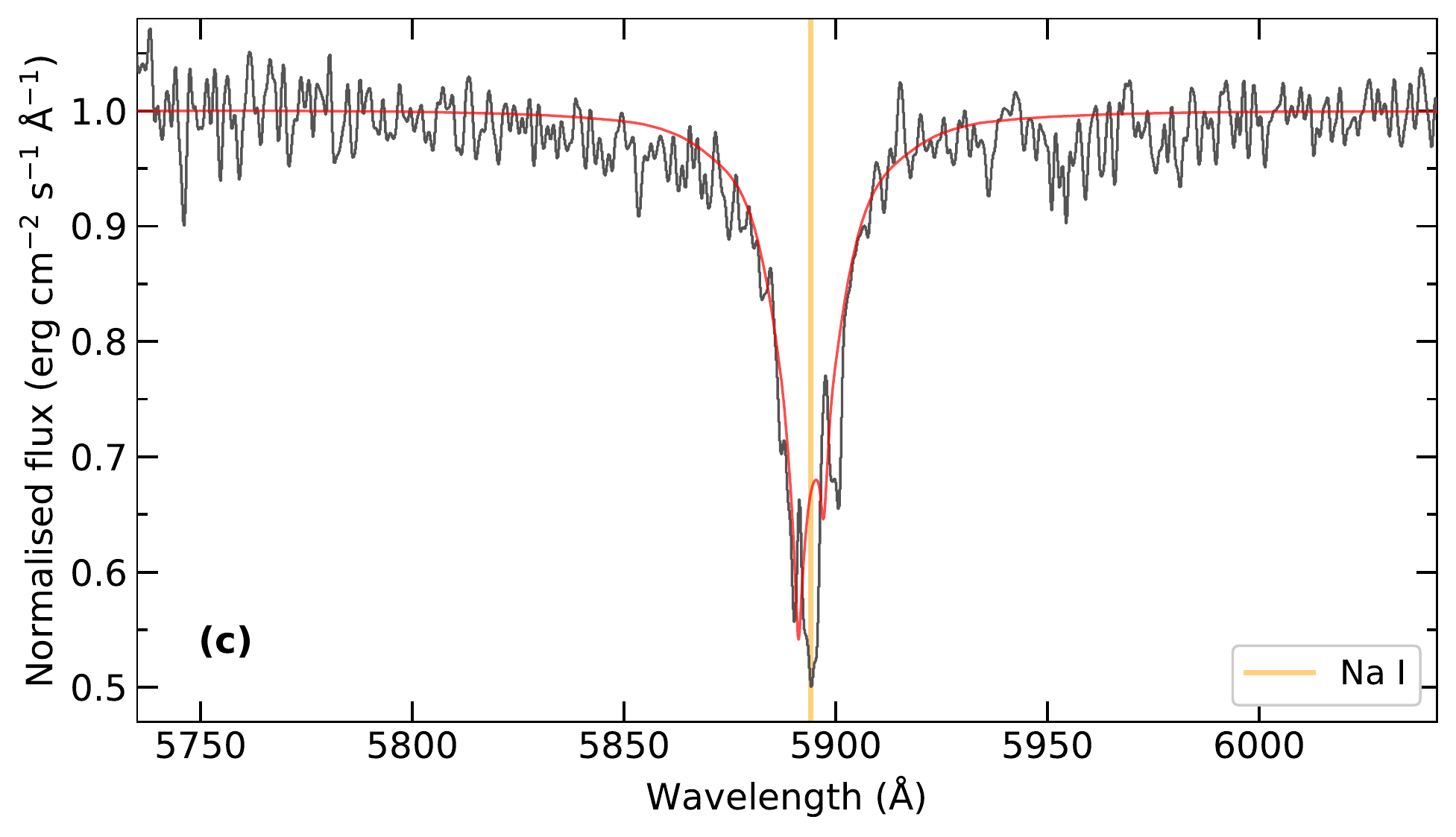}}
    \subfigure{\includegraphics[width=\columnwidth]{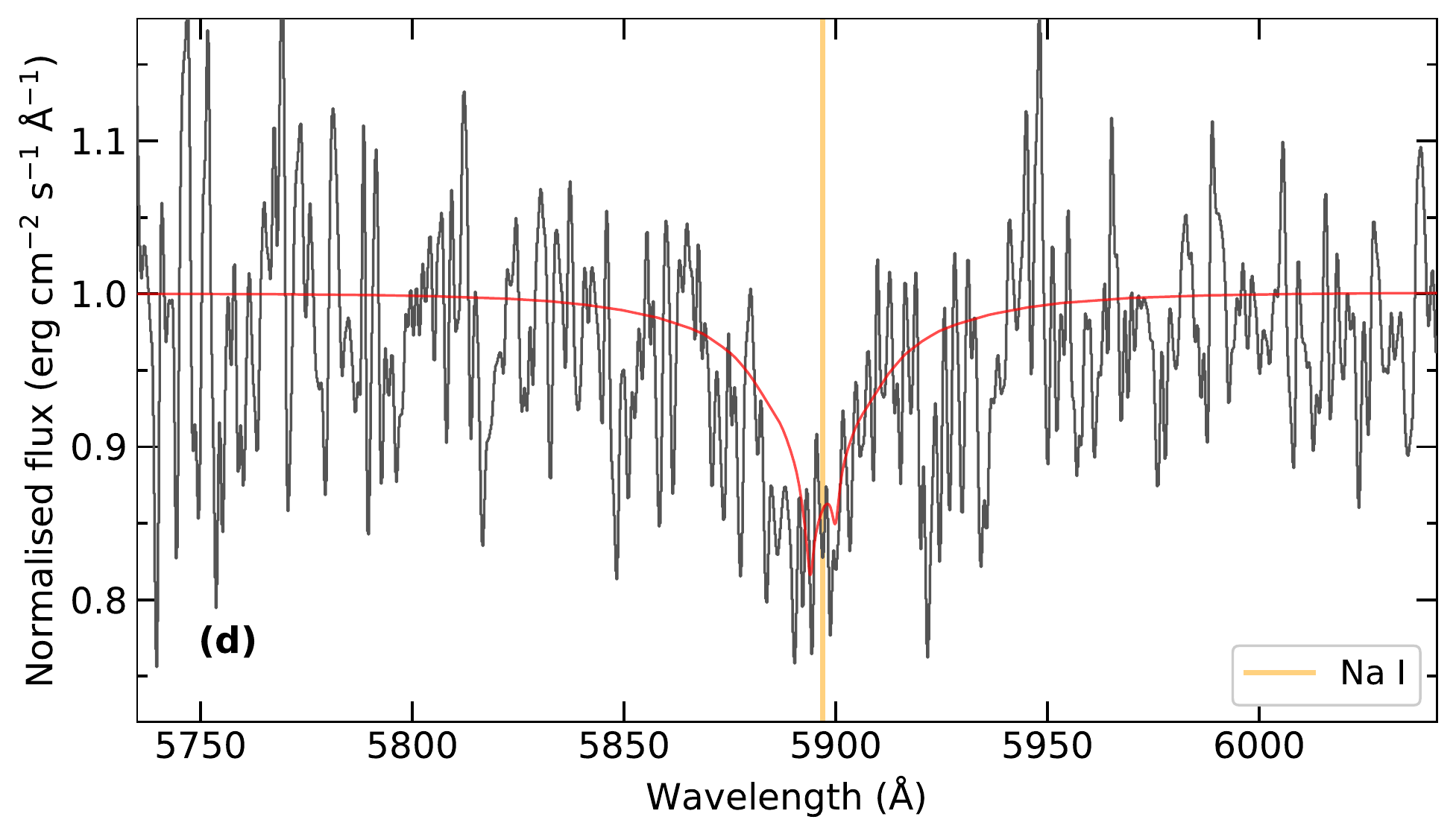}}
    \subfigure{\includegraphics[width=\columnwidth]{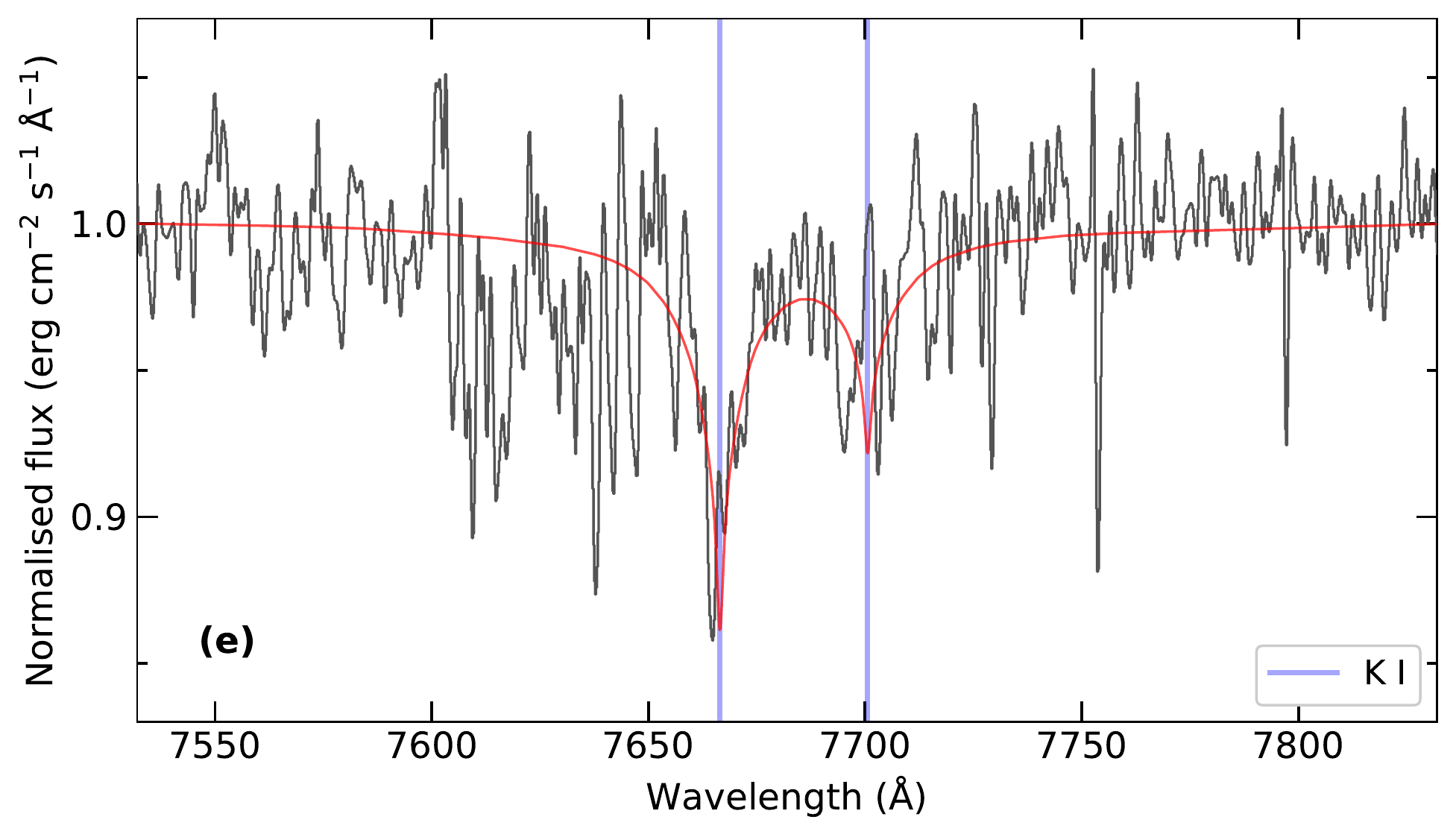}}
    \subfigure{\includegraphics[width=\columnwidth]{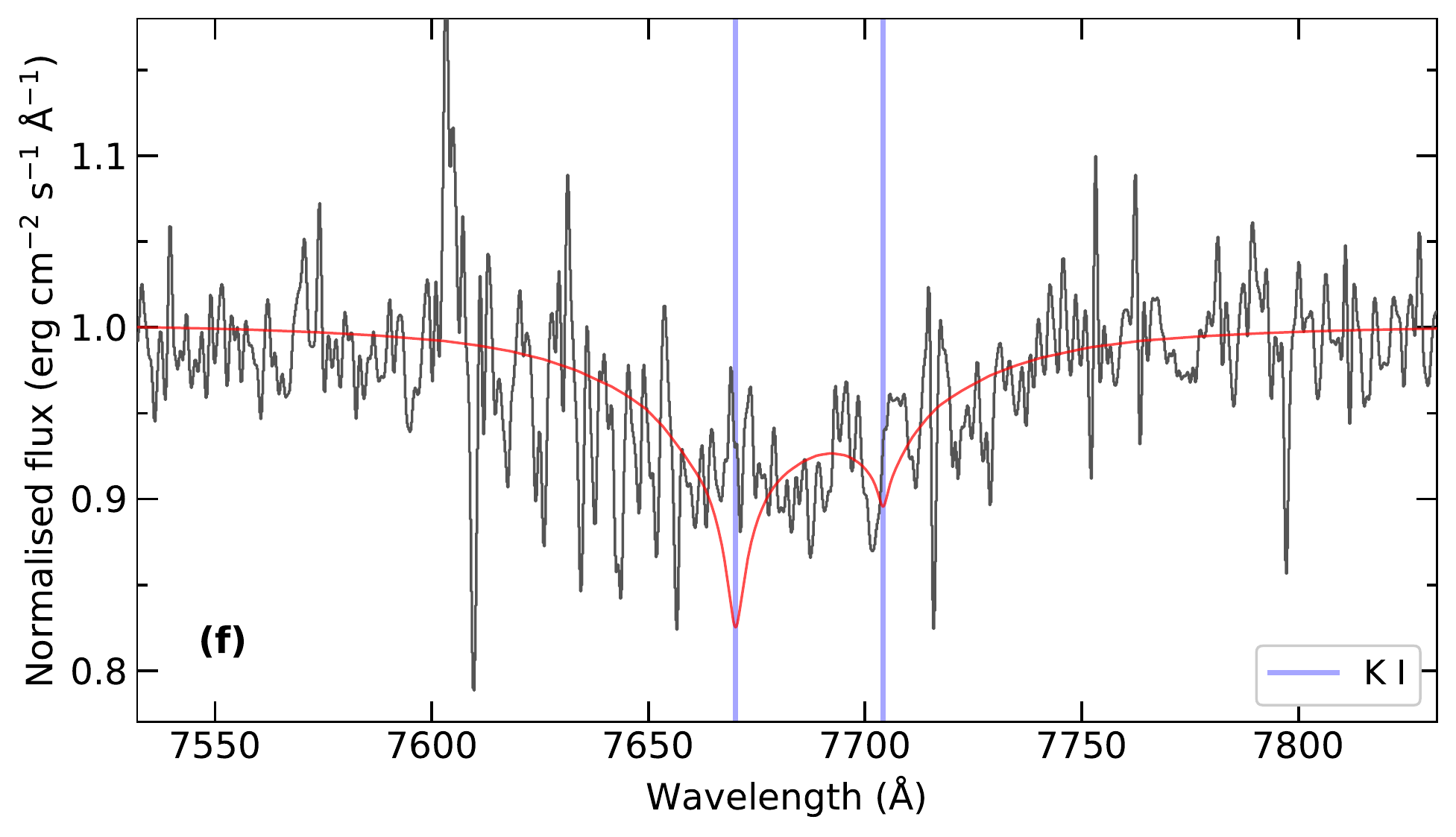}}
\caption{VLT X-Shooter normalised spectroscopic observations of WD\,J1922$+$0233 (left-hand panels) and WD\,J2147$-$4035 (right-hand panels) zoomed-in on metal detections. The coloured vertical bars indicate the detected metal lines at the appropriate radial velocity of each star (Section~\ref{sec:age_pop}). We have also applied the appropriate radial velocity to the best-fitting models, which are overlaid in red. The spectrum in panel (b) shows the Zeeman split \ion{Li}{i} (6708\,\AA) line in WD\,J2147$-$4035. All spectra and models are normalised to a continuum flux of one and convolved with a Gaussian with a FWHM of 1\,\AA\, for clarity.}
\label{fig:zoom-ins}
\end{figure*}

We used an iterative procedure to fit the observed spectra of WD\,J2147$-$4035 and WD\,J1922$+$0233, using model atmospheres including metals in the atmospheric structure calculations, with the best-fitting atmospheric parameters (\Teff, log~$g$, \logHHe; Table~\ref{tab:atmospheric}) derived in Section~\ref{sec:stellar_params} to constrain the abundances of detected metals in each star. The individual metal abundances in the models were iterated until a reduced $\chi^2 \approx 1$ was reached. Figure~\ref{fig:zoom-ins} shows the best-fitting model spectra for WD\,J2147$-$4035 and WD\,J1922$+$0233 in red, overlaid on the observed spectra which are zoomed-in on the the metal detections.

The metal abundance measurements of WD\,J2147$-$4035 and WD\,J1922$+$0233 are presented in Table~\ref{tab:abundance_ratios} and can be put into the context of metal abundances of other cool DZ white dwarfs found in the literature. The detection and subsequent abundance measurements of \ion{K}{i} were possible only after telluric line removal of the atmospheric O$_2$ absorption band near $\approx 7600$\,\AA\ was performed. However, we do not discount the possibility of the \ion{K}{i} line shape and strength being affected by the telluric removal as it is an imperfect process. We do not detect \ion{Li}{i} in WD\,J1922$+$0233 nor \ion{Ca}{i} in WD\,J2147$-$4035, so we constrained the observational upper limits of these metals by finding the abundance that produced the best-fitting model spectral line just within the noise level of the observed spectra. In Section~\ref{sec:atmos_abundance_ratios}, we discuss the accretion inferences made from the abundances of the detected metals and observational upper limits for both stars. 

C, O, Mg, Si, Al and Fe are not detected in WD\,J2147$-$4035 nor WD\,J1922$+$0233 but are commonly found in rocky exoplanetesimals and asteroids, and thus in the atmospheres of other DZ white dwarfs \citep{Zuckerman2007, Klein2010, Gansicke2012, Jura2014, Raddi2015}. Therefore, we also included these undetected metals in our model atmospheres and set them to Earth crust \citep{Rudnick2003} abundances relative to sodium (see Section~\ref{sec:atmos_abundance_ratios} for further explanation). However, these abundance ratios of Al and Fe for WD\,J1922$+$0233 produced absorption lines that were too strong by 0.64\,dex and 0.03\,dex, respectively, revealing their observational upper limits as $\log\textnormal{(Al/He)} < -12.65$ and $\log\textnormal{(Fe/He)} < -12.23$ which we used in the model instead. Table~\ref{tab:envelope params} displays the values used for all metals included in the model atmospheres of WD\,J2147$-$4035 and WD\,J1922$+$0233, including the measured abundances of detected metals, observational upper limits and the fixed abundances used for undetected metals.

\begin{table*}
	\centering
	\caption{Atmospheric abundance ratios for the white dwarfs in our DZ subsample. Abundances for WD\,J2147$-$4035 and WD\,J1922$+$0233 were derived in this work. We found the abundances for WD\,J1644$-$0449 in \citet{Kaiser2021}, WD\,J2356$-$2054 in \citet{Blouin2019}, and for WD\,J1824$+$1213, WD\,J1330$+$6435, WD\,J2317$+$1830 and WD\,J1214$-$0234 in \citet{Hollands2021}. Values for \logHHe\ are repeated here for clarity (also found in Table~\ref{tab:atmospheric}). Observational upper limits are given for undetected metals. Abundance ratios for Solar System benchmarks are calculated from values in \citet{mcdonough2000} for bulk Earth, \citet{Rudnick2003} for the continental crust, and \citet{Lodders2003} for CI chondrites and solar. Fields are left empty where no abundances were given in the literature.}
	\label{tab:abundance_ratios}
	\resizebox{\textwidth}{!}{\begin{tabular}{lcccccccccc} 
		\hline
		\hline
		& \logHHe & $\log\textnormal{(Ca/He)}$ & $\log\textnormal{(Na/He)}$ & $\log\textnormal{(Li/He)}$ & $\log\textnormal{(K/He)}$ & $\log\textnormal{(Ca/Na)}$ & $\log\textnormal{(Li/Na)}$ & $\log\textnormal{(K/Na)}$\\
		\hline
		\hline
		WD\,J2147$-$4035 & < $-$5.66 & < $-$13.20 & $-$13.10 $\pm$ 0.20 & $-$13.20 $\pm$ 0.20 & $-$12.90 $\pm$ 0.20 & < $-$0.10 & $-$0.10 $\pm$ 0.28 & 0.20 $\pm$ 0.28\\
		WD\,J1922$+$0233 & $-$2.69 $\pm$ 0.17 & $-$13.96 $\pm$ 0.20 & $-$12.60 $\pm$ 0.20 & < $-$13.60 & $-$13.10 $\pm$ 0.20 & $-$1.36 $\pm$ 0.28 & < $-$1.00 & $-$0.50 $\pm$ 0.28\\
		\hline
		WD\,J1824$+$1213 & $-$0.07 $\pm$ 0.10 & $-$10.28 $\pm$ 0.14 & $-$10.19 $\pm$ 0.07 & $-$11.95 $\pm$ 0.08 & -- & $-$0.21 $\pm$ 0.16 & $-$1.76 $\pm$ 0.11 & --\\
		WD\,J1330$+$6435 & < $-$4.00 & $-$10.94 $\pm$ 0.36 & $-$10.35 $\pm$ 0.12 & $-$11.96 $\pm$ 0.29 & -- & $-$0.59 $\pm$ 0.38 & $-$1.61 $\pm$ 0.31 & -- \\
		WD\,J1644$-$0449 & < $-$2.00 &  $-$9.5 $\pm$ 0.20 & $-$9.5 $\pm$ 0.20 & $-$11.2 $\pm$ 0.20 & $-$10.9 $\pm$ 0.20 & 0.00 $\pm$ 0.28 & $-$1.70 $\pm$ 0.28 & $-$1.40 $\pm$ 0.28\\
		WD\,J2356$-$2054 & $-$1.5 $\pm$ 0.2 & $-$9.4 $\pm$ 0.1$^b$ & $-$8.3 $\pm$ 0.2 & < $-$11.7$^c$ & < $-$10.4$^c$ & $-$1.10 $\pm$ 0.22 & < $-$3.40 & < $-$2.10\\
		WD\,J2317$+$1830 & $-$0.09 $\pm$ 0.09 & $-$10.79 $\pm$ 0.12 & $-$9.96 $\pm$ 0.07 & $-$11.19 $\pm$ 0.08 & < $-$10.50 & $-$0.83 $\pm$ 0.14 & $-$1.23 $\pm$ 0.11 & < $-$0.54\\
		WD\,J1214$-$0234 & $-$3.19 $\pm$ 0.18 & $-$10.08 $\pm$ 0.11 & $-$9.53 $\pm$ 0.06 & $-$11.83 $\pm$ 0.08 & $-$10.16 $\pm$ 0.08 & $-$0.55 $\pm$ 0.13 & $-$2.30 $\pm$ 0.10 & $-$0.63 $\pm$ 0.10\\
		\hline
		Bulk Earth & -- & -- & -- & -- & -- & 0.74 & $-$2.69 & $-$1.28 \\
		Continental crust & -- & -- & -- & -- & -- & $-$0.22 & $-$2.48 & $-$0.25 \\
		CI chondrites & 6.96 $\pm$ 0.05 & 5.00 $\pm$ 0.03 & 4.98 $\pm$ 0.03 & 1.96 $\pm$ 0.06 & 3.77 $\pm$ 0.05 & 0.02 $\pm$ 0.04 & $-$3.02 $\pm$ 0.07 & $-$1.21 $\pm$ 0.06 \\
		Solar & 1.10 $\pm$ 0.01 & $-$4.54 $\pm$ 0.02 & $-$4.60 $\pm$ 0.03 & $-$9.80 $\pm$ 0.10 & $-$5.78 $\pm$ 0.05 & 0.06 $\pm$ 0.04 & $-$5.20 $\pm$ 0.10 & $-$1.18 $\pm$ 0.06 \\
		\hline
	\end{tabular}}\\
\raggedright\footnotesize{
$^b$Updated value from \citet{Blouin2019_Ca}.\\
$^c$Updated value from \citet{Kaiser2021}.}
\end{table*}

\subsection{Magnetic nature of \texorpdfstring{WD\,J2147$-$4035}{WDJ2147--4035}}
\label{sec:Magnetic}
The cool magnetic white dwarf WD\,J1214$-$0234 was detected to have the first Zeeman split lithium line in \citet{Hollands2021}, where the magnetic field was measured to be 2.1\,MG. The \ion{Li}{i} spectral line in the X-Shooter spectrum of WD\,J2147$-$4035 indicates this star is also magnetic due to it being split into three components from Zeeman splitting. We found a best-fitting magnetic field strength of $0.55\pm0.03$\,MG as a result of fitting a constant magnetic field to the Zeeman split \ion{Li}{i} line; the three line components produced from this magnetic field are shown in Figure~\ref{fig:Li split line} by the vertical dashed pink lines. We do not analyse sodium or potassium spectral line splitting in WD\,J2147$-$4035 because these elements have much larger fine-structure splitting than lithium, hence the pattern is much more complicated than the triplet observed with the lithium line.

\subsection{A 13 h photometric period}
\begin{figure}
    \centering
    \includegraphics[width=\columnwidth]{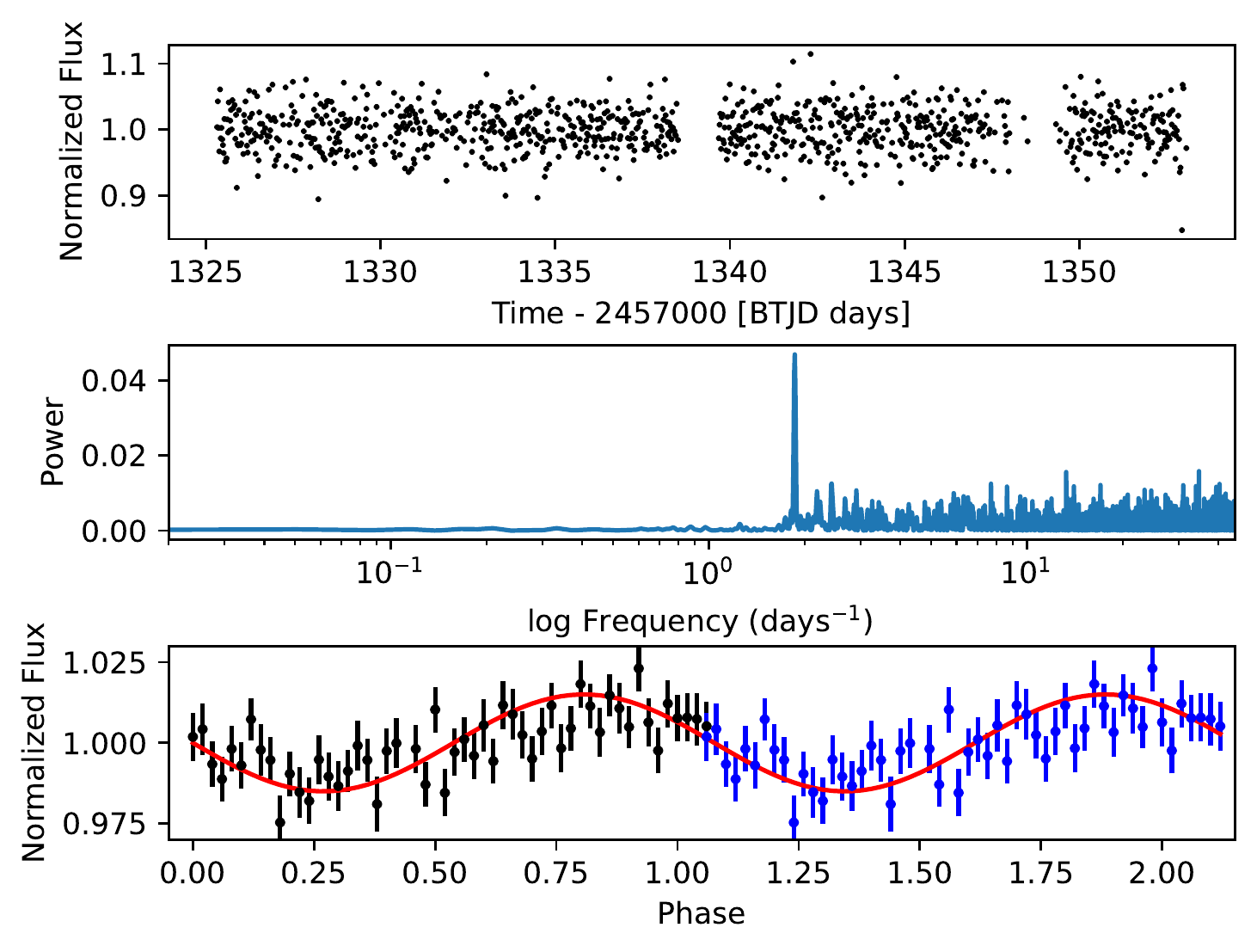}
\caption{FFI TESS data from the Sector one observation of WD\,J2147$-$4035. The cleaned, de-trended lightcurve is shown in the top panel. The middle panel shows the Lomb-Scargle periodogram, which exhibits a clear signal at $\simeq 0.54$\,days. The bottom panel displays the binned phase-folded lightcurve at this nominal period with the data repeated for illustrative purposes (blue points). A sine wave at the same period is shown in red overlay.}
\label{fig:WDJ2147_variability}
\end{figure}

Magnetism can result in inhomogeneous brightness distributions across the surface of the white dwarf, which in turn leads to photometric variability on the white dwarf spin period \citep[e.g.][]{Brinkworth2013}. We investigated the photometric variability of WD\,J2147$-$4035 by inspecting the full-frame images (FFI) of the Transiting Exoplanet Survey Satellite \citep[TESS;][]{TESS2014}, which observed the star in sectors one and 28. We split the TESS full frame image light curves into two segments per sector, before and after the gap occurring during the data downlink. We clipped $\simeq$\,three days at the end of each segment of the Cycle\,28 data, where the light curves displayed large-amplitude structures that most likely result from the data reduction (given that WD\,J2147$-$4035 has a TESS magnitude of $T\simeq 18.5$, extracting a light curve is challenging).  We then computed discrete Fourier transforms for the four individual light curves using the \textsc{TSA} context within \textsc{MIDAS} (Figure~\ref{fig:WDJ2147_variability}). All four periodograms contain a strong signal at $\simeq1.85\,\mathrm{d}^{-1}$, corresponding to a period of $\simeq13$\,h. We determined periods of $12.960\pm0.044$, $13.011\pm0.055$, $12.986\pm0.067$, $12.851\pm0.064$\,h from the four segments by fitting a sine curve to the data, with amplitudes of  $1.5 \pm 0.1$ per cent. We suggest that this consistently detected photometric signal represents the spin period of the white dwarf, which is compatible both in amplitude of the modulation and in period with the rotation variability commonly observed in magnetic white dwarfs.

\subsection{Carbon detection in \texorpdfstring{WD\,J2147$-$4035}{WDJ2147-4035}}
We made a tentative detection of C$_2$ in WD\,J2147$-$4035 due to the very broad, rounded absorption features centred at $\approx 4570$\,\AA, 5000\,\AA\ and 5400\,\AA\ being consistent with the three strongest C$_2$ Swan band systems \citep[progressions; ][]{Johnson1927} when broadened and empirically blueshifted from their nominal wavelengths. These absorption features are similar to those found in cool DQpec white dwarfs, which have distorted carbon features in their optical spectra presenting rounded profiles and centroid wavelengths blueshifted by $\approx 100-300$\,\AA\ \citep{Hall2008}. Studies have found that the high atmospheric pressure in cool DQpec stars drives this distortion \citep{Kowalski2010_DQpec}, however it is thought that a more complex combination of high atmospheric pressure, chemical composition and magnetic field strength is responsible for the varying distortions observed in individual stars, but the specific mechanisms are currently not fully understood \citep{Liebert1978, Bergeron1994, Schmidt1995, Schmidt1999, Hall2008, Blouin2019d}. 

One DQpecP and four DQpec stars are found in the MWDD with $\Teff < 5000$\,K (in increasing temperature order): LP\,351$-$42, SDSS\,J1803202.57+232043.3, SDSS\,J082955.77+183532.6, GJ\,3614 and PM\,J12476+0646. LP\,351$-$42 and GJ\,3614 have strong magnetic fields of 100\,MG and 50\,MG, respectively, in addition to LP\,351$-$42 having detectable polarisation hence its classification of DQpecP \citep{McCook_Sion1999}. The other three white dwarfs do not have detected magnetic fields and none of the stars have published \logHHe\ atmospheric abundances. The C$_2$ progressions in all five white dwarfs are blueshifted with respect to the nominal wavelengths, with the largest distortions evident in GJ\,$3614$ and LP\,351$-$42. The C$_2$ progressions observed in LP\,351$-$42 are in excellent agreement with the broad absorption features in WD\,J2147$-$4035 and are shown in Figure~\ref{fig:VLT_spectra} with aqua vertical lines. LP\,351$-$42 is $\approx 1250$\,K warmer than WD\,J2147$-$4035 and has a significantly stronger magnetic field so the reason for this agreement is unclear. An empirical blueshift of $\approx 20$\,\AA\ to the progressions in SDSS\,J1803202.57$+$232043.3, which is $\approx 1350$\,K warmer than WD\,J2147$-$4035, also provides excellent agreement to the absorption features in WD\,J2147$-$4035. The other three DQpec stars require larger blueshifts to their observed progressions to fit the features of WD\,J2147$-$4035.

We do not attempt to fit the distorted C$_2$ progressions in WD\,J2147$-$4035 using our model atmospheres as the high atmospheric density, relatively weak bands and low signal-to-noise of the X-Shooter spectrum make a quantitative analysis particularly challenging. However, we still assign the spectral type DZQH to this star. The combination of metals polluting WD\,J2147$-$4035, including sodium, lithium, potassium and tentatively carbon, in addition to its ultra-cool temperature and magnetic nature, makes this an extremely rare white dwarf. 

\section{Discussion}
\label{sec:Discussion}

\subsection{Neutral line broadening}
\label{sec:Neutral line broadening}
\citet{Hollands2021} used an empirically determined factor of ten to reduce the neutral-broadening constant of lithium in their models to extract a good fit to the observations. We found that a reduction factor of 100 gave the best-fitting model to the equivalent width and shapes of observed absorption lines in WD\,J2147$-$4035 and WD\,J1922$+$0233. Our ad-hoc treatment of the neutral-broadening constant was applied for all observed metals. 

The standard neutral-broadening constants for WD\,J2147$-$4035 and WD\,J1922$+$0233 were calculated using the impact approximation, which is the standard theory of line broadening \citep{Kolb_Griem1958, Griem1960, Griem1974}. The condition for this application is that the time of effective interaction between two particles, deemed the emitter and perturber, is much shorter than the time between interactions. In our model, the mean distance between emitter and perturber in the photosphere is $\approx 1$\,\AA\, therefore this is definitely not the case -- the emitter is constantly affected by interactions with the perturbers. Alternative approaches include the quasistatic theory \citep{Mozer_Baranger1960, Baranger1962}, which traditionally assumes van der Waals interactions, or the unified theories \citep{Voslamber1969, Smith1969}. However, none of these are applicable at $\approx 1$\,\AA\, distances.

A study performed by \citet{Nur2015} with high density argon corona plasma at very low temperatures found that with increasing hydrostatic pressure the line broadening of gaseous \ion{Ar}{i} follows a positive linear relation with particle density, yet when transitioning to a fluid this relation ceases and the line width appears significantly narrower than expected from extrapolation and even decreases at higher densities (see Fig. 5 of \citealt{Nur2015}). \citet{Nur2015} used a maximum particle density an order of magnitude lower than the density in the photosphere of WD\,J2147$-$4035 and WD\,J1922$+$0233, so the use of an empirical broadening constant on metal lines within models for ultra-cool DZ white dwarfs may have a plausible justification. Even so, the dense fluid physics involved within the extremely high-pressure photospheres of these stars requires further study so an appropriate theory of line broadening at these densities can be created.

\subsection{Nature of \texorpdfstring{WDJ\,2147$-$4035}{WDJ2147-4035} and \texorpdfstring{WDJ\,1922$+$0233}{WDJ1922+0233}}
\label{sec:nature}
The observations of WD\,J2147$-$4035 suggest much milder atmospheric CIA compared to WD\,J1922$+$0233 because it is very red, with $g-z = 2.29$\,mag. There could be two overlapping explanations for the nature of WD\,J2147$-$4035: a vastly different \logHHe\,ratio to WD\,J1922$+$0233, resulting in a much lower H$_2$ abundance and CIA opacity; or a significant change in non-ideal effects, resulting in the dissociation of H$_2$ molecules. The presence of metal lines is helpful to break this degeneracy, although the empirical broadening factor needed to fit them in both objects is a significant impairment to extract atmospheric conditions from the metal lines. The main evidence that can be gathered from the metal lines is that hydrogen-dominated atmospheres can be excluded for both objects, making a strong case that cool white dwarfs with unusually strong CIA have mixed \logHHe\,atmospheres, as previously accepted from indirect evidence \citep{Kilic2020}. Furthermore, strong non-ideal effects that could ionise more H$_2$ and therefore force the model fluxes to redder wavelengths, would also ionise alkali metals with low ionisation potentials compared to the  H$_2$ molecule. Therefore, metal line observations also put an upper limit on the strength of non-ideal effects.

The unusual position of WD\,J1922$+$0233 on the HRD (Figure~\ref{fig:HR}) suggests its nature can be explained by two scenarios: either it is relatively massive which caused it to have a higher absolute magnitude than stars of more moderate mass, hence it is placed vertically below the main white dwarf cooling sequence; or, flux suppression in the red optical and IR due to CIA caused the star to have a peak flux emission at bluer wavelengths, resulting in a horizontal placement blueward of the main white dwarf cooling sequence. We conclude the latter is true as we derived a moderate mass for WD\,J1922$+$0233 in  Section~\ref{sec:stellar_params} and found its SED is indicative of strong CIA opacity due to an extreme IR flux deficit in Section~\ref{sec:Spectroscopic_observations}. Also, cool white dwarfs typically have dense photospheres compared to their warmer counterparts; however, an increasingly higher atmospheric composition of \logHHe\,decreases the density. The complex interplay between photospheric density, non-ideal EOS effects and the abundance of H$_2$ controls the intensity of CIA \citep{Blouin2017,Blouin2018a}. The maximum intensity of the H$_2$-He CIA corresponds to an atmospheric abundance of $\logHHe \simeq -3$ \citep{Kilic2020}, which is close to our best-fitting abundance of $\logHHe = -2.69 \pm 0.17$ for WD\,J1922$+$0233. This is consistent with observations of strong CIA opacity in the relatively blue WD\,J1922$+$0233. 

Given available information, the best explanation for the colour differences between WD\,J2147$-$4035 and WD\,J1922$+$0233 is that the former has a much lower \logHHe\ ratio and only mildly enhanced non-ideal effects compared to WD\,J1922$+$0233, resulting in much milder CIA.

\subsection{Nature of IR-faint white dwarfs}
\label{sec:IR-faint}
We note that our grid of models is able to provide reasonable solutions for the spectroscopy and photometry of both WD\,J2147$-$4035 and WD\,J1922$+$0233 with the same microphysics, despite their vastly different positions in the HRD. Yet, tests using our grid of models with $3000 < \Teff < 10\,000$\,K, $7.0 < \log g < 9.5$ and $-5.0 < \logHHe < 0.0$ proved unable to explain the ultra-blue sequence of DC white dwarfs observed in Figure\,\ref{fig:HR} and previously identified in \citet{Kilic2020, Bergeron2022}. The most likely explanation for this behaviour is that the intricate balance between CIA and non-ideal effects is not yet fully accurate in our models for these stars.

Nevertheless, the discovery of WD\,J1922$+$0233, the first IR-faint DZ white dwarf with flux suppression in the red optical and IR from strong CIA, and its comparison with other ultra-cool DZ stars, opens a new window to understand the atmospheric composition of IR-faint white dwarfs. These observations suggest that they form a small sub-sample of all cool stellar remnants that have a rather narrow range of \logHHe\ around $\approx -3.0$, with the exact value subject to significant modelling uncertainties. This is supported by the finding that all other ultra-cool DZ white dwarfs, with a \logHHe\ ratio that is either small ($<-6.0$) or large ($>-1$) according to Table~\ref{tab:atmospheric}, reside closer to the main cooling sequence in the various HRDs. WD\,J1214$-$0234 has $\logHHe \approx -3.0$ so could be a future member of the ultra-blue sequence, as its current \Teff\ is too high to allow this transition. 

The spectral evolution of white dwarfs, i.e. the evolution of the ratio of He- to H-dominated atmospheres, has been extensively studied in recent years \citep{Blouin2019d,Cunningham2020,McCleery2020,Bedard2020,Carlos2022}. These studies suggest that the ratio of He- to H-dominated atmospheres is $\approx$ 0.3 in the range $5000 < \Teff < 6000$\,K, where it is possible to constrain atmospheric composition from the presence or absence of the H$\alpha$ line. However, \citet{GF2020} have demonstrated from optical and IR HRDs that the vast majority of cool ($\Teff < 5000$\,K) DC white dwarfs, residing near the main cooling sequence, could have either H- or He-rich atmospheres. In other words, there is no clear separation in colour space between these two atmospheric classes, preventing any robust constrain on the spectral evolution of very cool white dwarfs. This is consistent with both He- and H-rich ultra-cool DZ white dwarfs residing at the end of the main white dwarf cooling track in the different HRDs. We disagree with the suggestion of \citet{Kilic2020,Bergeron2022} that the observations provide evidence of spectral evolution towards strong H-atmosphere dominance for cool DC white dwarfs ($\Teff < 5000$\,K).

\subsection{Total age, mass and population membership}
\label{sec:age_pop}

\begin{figure}
    \centering
    \subfigure[]{\includegraphics[width=\columnwidth]{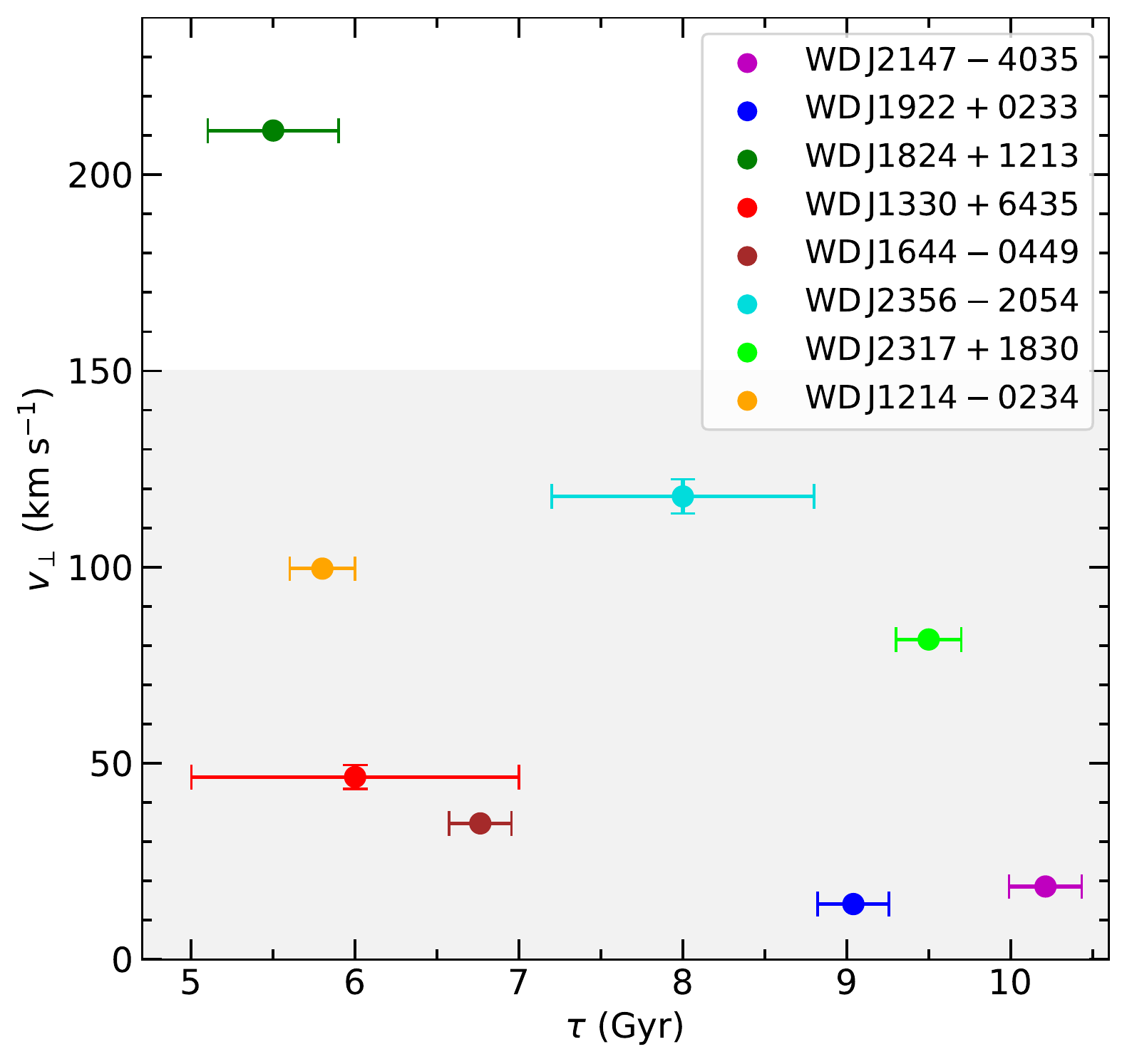}\label{fig:coolingAge_vtan}}
    \subfigure[]{\includegraphics[width=\columnwidth]{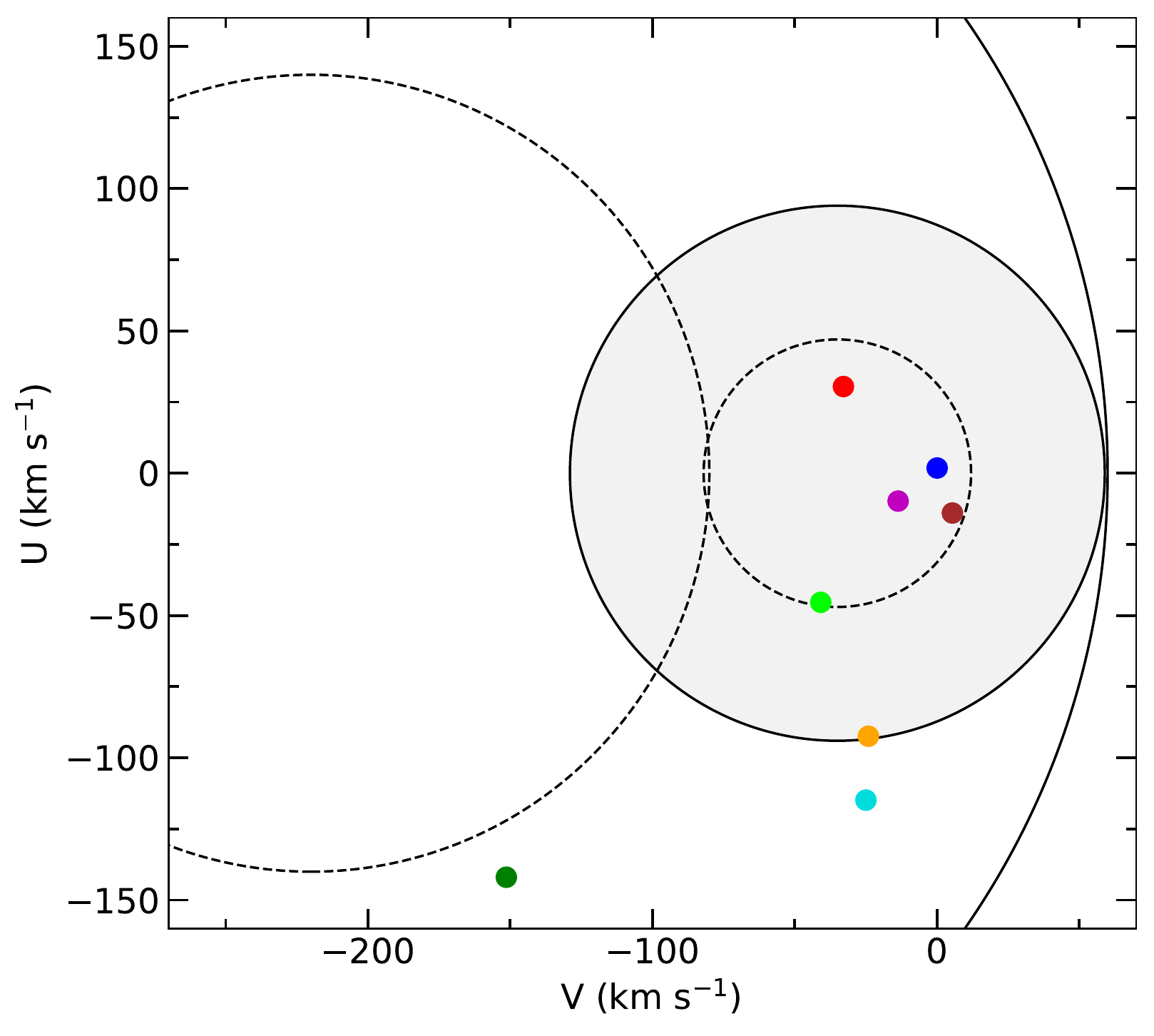}\label{fig:V_U_components}}
\caption{Kinematics of each DZ white dwarf in our subsample. Tangential velocity ($v_{\perp}$) as a function of cooling age ($\tau$) is displayed in (a), where error bars represent $1\sigma$ uncertainties but are omitted when they lie within the points. The shaded region represents $v_{\perp}$ consistent with Galactic disc membership. The $v_{\perp}$ of each subsample DZ star is transformed into Galactic radial ($U$) and rotational ($V$) velocity components assuming zero radial velocity; we display these velocity dispersions in (b). The $1\sigma$ and $2\sigma$ velocity dispersions of the Galaxy's halo and disc (shaded) are marked by dashed and solid ellipses, respectively. The legend apply to both panels.}
\label{fig:age_pop}
\end{figure}

Using the $M_{\mathrm{WD}}$ we derived (Table~\ref{tab:atmospheric}) and the initial-to-final-mass relation (IFMR) from \citet{Cummings2018}, we calculated the progenitor mass of WD\,J2147$-$4035 to be $2.47\pm0.22$ M$_\odot$ with a main-sequence lifetime of 0.5\,Gyr, with asymmetric $1\sigma$ error bars ranging from $0.4 - 0.7$\,Gyr. We therefore found the total age to be $10.7\pm0.3$\,Gyr which is $\approx 1$\,Gyr above WD\,J2317$+$1830 \citep{Hollands2021}, the most massive object in our DZ subsample (Table~\ref{tab:atmospheric}).

Following the same procedure as for WD\,J2147$-$4035, we calculated an approximately solar progenitor mass of $1.01\pm0.40$\,M$_\odot$ for WD\,J1922$+$0233. The main-sequence lifetime of WD\,J1922$+$0233 for solar metallicity therefore has a median of 10.5\,Gyr, ranging from 3.8\,Gyr to surpassing the age of the universe \citep{Hurley2000}. Together with the cooling age (Table~\ref{tab:atmospheric}), we found the total age of WD\,J1922$+$0233 to be over 13.8\,Gyr which is incompatible with the age of the Milky Way. We conclude that our current model underestimates the mass of WD\,J1922$+$0233, likely due to the physics of strong CIA not being fully understood and the uncertainties from our fits only being of a statistical nature, hence its total age remains unknown.

The $M_{\mathrm{WD}}$ of remnants with $\Teff < 5000$\,K are systematically lower\footnote{The average $M_{\mathrm{WD}}$ for the local 40\,pc white dwarf volume sample is 0.66\,M$_{\odot}$ for $\Teff > 5000$\,K and 0.52\,M$_{\odot}$ for $\Teff < 5000$\,K \citep{GF2021}.} than those of warmer white dwarfs \citep{Bergeron2019, McCleery2020, Tremblay2020}, which is inconsistent with the Galactic model prediction of white dwarfs cooling at constant mass \citep{Tremblay2016}. This is also the case where strong CIA is present, and likely associated with the physical uncertainties that render their atmospheres notoriously challenging to model \citep[e.g.][and see Sections~\ref{sec:model-atmospheres} and \ref{sec:Neutral line broadening}]{Gianninas2015, Bergeron2022}.

The stars in our DZ subsample also suggest $M_{\mathrm{WD}}$ underestimations. Three subsample stars have $M_{\mathrm{WD}} \lesssim 0.51$\Msun\ which is not possible from single stellar evolution so these masses must be underestimated \citep{Cummings2018}. Combined with the uncertainties from our fits and existing published ones being of a statistical nature so the real uncertainties are likely larger, it is probable the derived masses of our DZ subsample are low estimates. Although, even if the $M_{\mathrm{WD}}$ of WD\,J2147$-$4035 was underestimated by $\approx 0.1$\,\Msun, this would only make the total age smaller by $\approx 0.5$\,Gyr, given the large cooling age and short main-sequence lifetime. Therefore, WD\,J2147$-$4035 is unambiguously one of the oldest known metal-polluted white dwarfs.

To understand the population membership of the DZ subsample, we calculated the tangential velocity ($v_{\perp}$) of each white dwarf using the method described in \citet{McCleery2020} and show the results in Table~\ref{tab:astrometry} and Figure~\ref{fig:coolingAge_vtan} as a function of cooling age ($\tau$). WD\,J2147$-$4035 has $v_{\perp} = 18.59\pm0.26$\,\kms\ and WD\,J1922$+$0233 has $v_{\perp} = 14.11\pm0.15$\,\kms. All but one remnant have $v_{\perp} < 150$\,\kms\ suggesting they are Galactic disc candidates \citep{Gianninas2015}. We do not differentiate between thick and thin disc membership in this work, we instead treat them as one Galactic disc population. WD\,J1824$+$1213 has a significantly higher $v_{\perp}$ than the other objects and, together with its relatively small $\tau$ of $5.5\pm0.4$\,Gyr, is our only Galactic halo candidate -- this conclusion was also drawn by \citet{Hollands2021} and \citet{Kilic2019}. Assuming that halo membership makes it the oldest object in the subsample, it requires a large main-sequence lifetime. This is consistent with WD\,J1824$+$1213 having the lowest $M_{\mathrm{WD}}$ in Table~\ref{tab:atmospheric}.

To further analyse the population membership of our DZ subsample, we transformed their $v_{\perp}$ into Galactic velocity components $U$, $V$ and $W$, which indicate motion radially away from the Galactic center, in the direction of the Galaxy's rotation and perpendicular to the disc, respectively, using \textit{Gaia} EDR3 astrometry and corrected parallaxes from Table~\ref{tab:astrometry}; we present the velocity component values there also. These velocity components are computed in \citet{Hollands2021} for WD\,J1824$+$1213, WD\,J1330$+$6435, WD\,J1214$-$0234 and WD\,J2317$+$1830 using \textit{Gaia} DR2 astrometry so updated measurements are reported here. We assumed zero radial velocity in our calculations following standard practices \citep{Hollands2021, Kaiser2021} and due to uncertainties in the gravitational redshift corrections from the mass uncertainties. We rely on the previous study of \citet{Oppenheimer2001_membership} to analyse the velocity dispersions of our subsample and assess their $1\sigma$ or $2\sigma$ disc or halo membership. We plot our results in Figure~\ref{fig:V_U_components}, where the largest dashed and solid ellipses encapsulate the velocity dispersions consistent with the Galaxy's halo up to $1\sigma$ and $2\sigma$, respectively. The shaded region represents velocity dispersions consistent with the Galaxy's disc up to $1\sigma$ (dashed ellipsis) and $2\sigma$ (solid ellipsis). WD\,J1824$+$1213 is a likely halo candidate as it has velocity components consistent with $2\sigma$ velocity dispersions of the Galaxy's halo. WD\,J2147$-$4035, WD\,J1922$+$0233, WD\,J1330$+$6435, WD\,J1644$-$0449 and WD\,J2317$+$1830 have velocity components consistent with at least $2\sigma$ of the Galaxy's disc hence likely have disc membership. The membership conclusions for the above five subsample objects are consistent with those drawn from their $v_{\perp}$. WD\,J2356$-$2054 and WD\,J1214$-$0234 both have an uncertain disc or halo membership allocation in $U$ vs. $V$ space, yet their $v_{\perp}$ indicates disc membership.

We measured the wavelength shift of the lithium spectral line central component in WD\,J2147$-$4035 and the sodium spectral line in WD\,J1922$+$0233 compared to rest wavelengths and calculated the line velocities of each star as $120.6 \pm 16.8$\,\kms\ and $-20.1 \pm 10.2$\,\kms, respectively. Despite uncertainties in the calculations, we corrected for each star’s gravitational redshift to determine radial velocities, $v_\mathrm{rad}$, of $80.6 \pm 17.1$\,\kms\ for WD\,J2147$-$4035 and $-49.1 \pm 10.6$\,\kms\ for WD\,J1922$+$0233. Accounting for $v_\mathrm{rad}$ in the calculations of $U$, $V$ and $W$ for WD\,J2147$-$4035 and WD\,J1922$+$0233, both stars still indicate Galactic disc membership.

We do not include the total ages of WD\,J2147$-$4035 and WD\,J1922$+$0233 in our membership assessment because previous estimates of the Galactic disc age from white dwarfs \citep{Winget1987, Kilic2017} are thought to be underestimated due to $^{22}$Ne dilution cooling delays \citep{Tremblay2019, Blouin2020, Kilic2020} being omitted from earlier crystallisation calculations. Nevertheless, both white dwarfs could be utilised in future studies for constraining an upper age limit for the disc of the Milky Way.

\subsection{Accreted planetary debris}
\label{sec:accreted debris}

\begin{table*}
	\centering
	\caption{For each metal (Z) included in the model atmospheres for WD\,J2147$-$4035 and WD\,J1922$+$0233, the first ten rows show the atmospheric abundances ($\log\textnormal{(Z/He)}$) we used (observationally detected metals and subsequently measured abundances include uncertainties; observational upper limits are denoted with a less than (<) symbol; undetected metals were fixed to the abundances shown and have no associated uncertainties), diffusion timescales ($\tau_{\mathrm{z}}$) in Myr and metal masses ($M_{\mathrm{z}}$) in $\times 10^{15}$\,g. The subsequent rows show the logarithm of the fractional convection zone mass ($M_{\mathrm{cvz}}/M_{\mathrm{WD}}$) and the minimum mass of the accreted parent body when considering only metals ($M_{\mathrm{z,tot}}$) and hydrogen ($M_{\mathrm{H}}$). Convective overshoot \citep{Cunningham2019} with a pressure scale height of one is included in the envelope code \citep{Koester2020} used to calculate these parameters.}
	\label{tab:envelope params}
	\begin{tabular}{lcccccc}
		\hline
		\hline
		& \multicolumn{3}{c}{\textbf{WD\,J2147$-$4035}} & \multicolumn{3}{c}{\textbf{WD\,J1922$+$0233}} \\
		Z & $\log\textnormal{(Z/He)}$ & $\tau_{\mathrm{z}}$ & $M_{\mathrm{z}}$ & $\log\textnormal{(Z/He)}$ & $\tau_{\mathrm{z}}$ & $M_{\mathrm{z}}$ \\
		\hline
		\hline
		Li & $-$13.20 $\pm$ 0.20 & 4.14 & 0.699 & < $-$13.60 & 9.21 & 0.55 \\
		C & $-$14.69 & 3.84 & 0.04 & $-$14.58 & 8.73 & 0.10 \\
		O & $-$11.63 & 2.76 & 59.94 & $-$11.38 & 6.38 & 209.34 \\
		Na & $-$13.10 $\pm$ 0.20 & 1.70 & 2.92 & $-$12.60 $\pm$ 0.20 & 4.01 & 18.11 \\
		Mg & $-$12.00 & 1.68 & 38.79 & $-$12.73 & 3.97 & 14.19 \\
		Al & $-$12.61 & 1.44 & 10.59 & < $-$12.65 & 3.43 & 18.92 \\
		Si & $-$12.10 & 1.44 & 35.63 & $-$11.61 & 3.44 & 216.20 \\
		K & $-$12.90 $\pm$ 0.20 & 0.97 & 7.87 & $-$13.10 $\pm$ 0.20 & 2.35 & 9.75 \\
		Ca & < $-$13.20 & 0.97 & 4.04 & $-$13.96 $\pm$ 0.20 & 2.35 & 1.38 \\
		Fe & $-$13.14 & 0.64 & 6.47 & < $-$12.23 & 1.57 & 103.01 \\
		\hline
		log($M_{\mathrm{cvz}}/M_{\mathrm{WD}}$) & \multicolumn{3}{c}{$-$5.32} & \multicolumn{3}{c}{$-$4.95} \\
		$M_{\mathrm{z,tot}}$ & \multicolumn{3}{c}{$1.67 \times 10^{17}$\,g} & \multicolumn{3}{c}{$5.92 \times 10^{17}$\,g} \\
		$M_{\mathrm{H}}$ & \multicolumn{3}{c}{$1.43 \times 10^{21}$\,g} & \multicolumn{3}{c}{$6.50 \times 10^{24}$\,g} \\
		\hline
	\end{tabular}
\end{table*}

\begin{figure}
    \centering
    \subfigure[]{\includegraphics[width=\columnwidth]{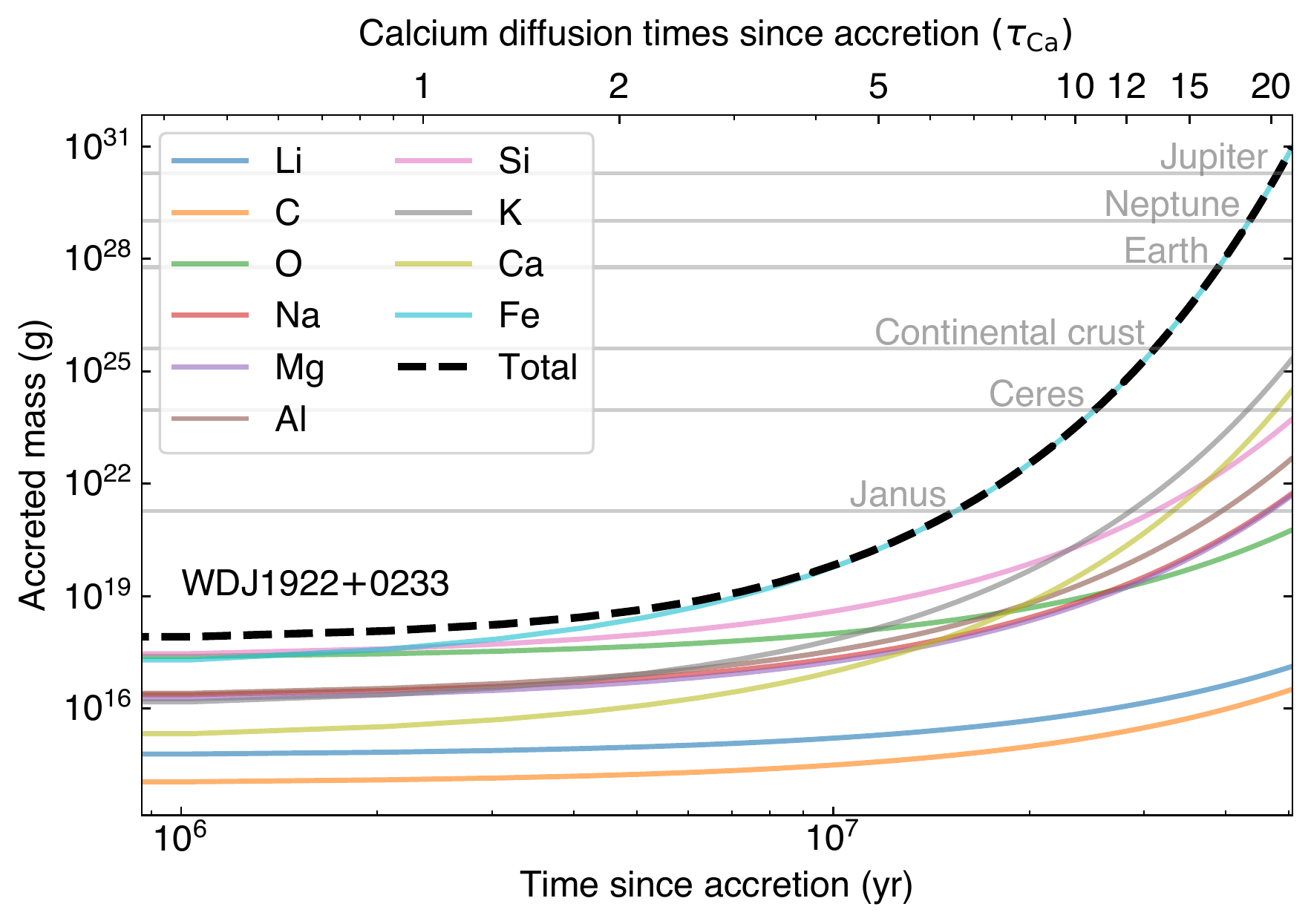}\label{fig:parent body mass WDJ1922}}
    \subfigure[]{\includegraphics[width=\columnwidth]{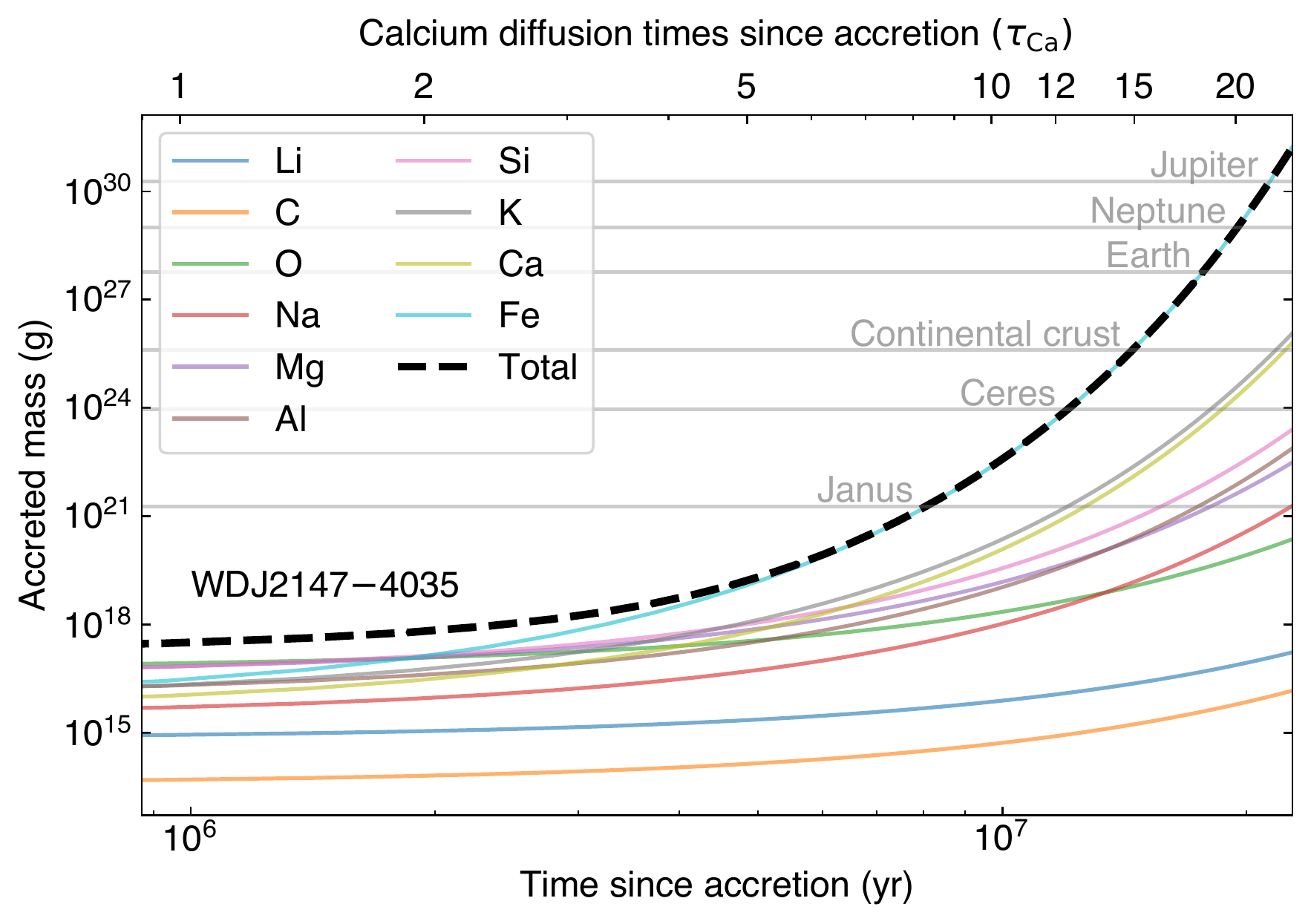}\label{fig:parent body mass WDJ2147}}
\caption{Masses of metals ($M_{\mathrm{z}}$) in the convection zone of (a) WD\,J1922$+$0233 and (b) WD\,J2147$-$4035, which are representative of the accreted parent body masses, as a function of time and calcium diffusion times ($\tau_\mathrm{{Ca}}$) since accretion ceased. The $M_{\mathrm{z}}$ of metals used in the model atmospheres are shown by coloured solid lines. The black dashed line represents the minimum accreted parent body mass ($M_{\mathrm{z,tot}}$) for each white dwarf depending when accretion stopped. We cannot unambiguously determine the fraction of hydrogen mass accreted from planetary debris from the total mass of hydrogen in the CVZ ($M_{\mathrm{H}}$), therefore the mass of accreted hydrogen is not considered in $M_{\mathrm{z,tot}}$. The mass of six Solar System objects are shown for comparison.}
\label{fig:parent body mass}
\end{figure}

Relative metal abundances measured in the atmospheres of metal-polluted white dwarfs using spectral fitting (Section~\ref{sec:Spectroscopic_observations}) inform us of the bulk compositions of debris accreted from planetary systems. However, atmospheric abundance ratios change over time due to atomic diffusion causing metals to settle from the bottom of the convection zone (CVZ) to the interior. The abundance evolution history of DZ stars can be traced backwards in time by taking into account the diffusion (sinking) timescales ($\tau_{\mathrm{z}}$) of individual elements, which each have different rates of diffusion. Cool DZ stars have long $\tau_{\mathrm{z}}$ which is the primary mechanism responsible for the long-term evolution of atmospheric abundances, yet the $\tau_{\mathrm{z}}$ of all elements are many orders of magnitude shorter than white dwarf cooling ages \citep{Koester2009}. Therefore, the observation of metal lines in the atmospheres of white dwarfs requires recent or active accretion \citep{Vauclair1979}. 

Accretion episodes are typically described as three simplified phases \citep{Koester2009}: the increasing state, where material is actively accreted into the white dwarf photosphere so its atmospheric composition initially resembles that of the debris as it is instantaneously mixed throughout the CVZ, then the composition diverges as metals diffuse at their individual diffusion velocities through the bottom of the CVZ; steady-state, or accretion-diffusion equilibrium, which is reached if the duration of the accretion event extends over several diffusion timescales; and the decreasing state, where accretion has stopped and the atmospheric abundances exponentially decrease according to the individual diffusion time scales of each metal.

Based on our abundance analysis (see Section~\ref{sec:atmos_abundance_ratios}), we assume that WD\,J2147$-$4035 and WD\,J1922$+$0233 are not currently accreting but have instead retained debris from an earlier accretion event. Therefore, we proceed with our analysis on the assumption that both white dwarfs have stopped accreting (i.e. are in the decreasing phase).

A lower limit of the accreted parent body mass ($M_{\mathrm{z,tot}}$) can be constrained by the sum over all individual metal masses ($M_{\mathrm{z}}$) currently contained in the CVZ, as this measurement is independent of the accretion state \citep{Izquierdo2021}. We used the envelope code described in \citet{Koester2020} to compute $\tau_{\mathrm{z}}$, the fractional CVZ mass ($M_{\mathrm{cvz}}/M_{\mathrm{WD}}$) and the mass fraction of each metal included in the model atmospheres of WD\,J2147$-$4035 and WD\,J1922$+$0233 according to the photospheric abundance, and included convective overshoot \citep{Cunningham2019} with a pressure scale height ($H_{\rm p}$) of one. Details on the physics and methods used in the envelope code can be found in \citet{Koester2020}, but most important to note for this work is that the boundary conditions for the envelope integration were taken from the atmosphere models at a Rosseland optical depth of $\approx 300$. At this depth, the ionisation is significantly higher than in the photosphere ($\tau \approx 2/3$) and CIA opacity is much less important. Therefore the diffusion timescales and fluxes in our final model atmosphere, as well as for our atmosphere model without any CIA absorption, differed very little. Uncertainties in the CIA treatment were hence negligible for these diffusion data.

The resulting parameters computed with our envelope code are listed in Table~\ref{tab:envelope params}. For metals with observationally determined upper limits (see Table~\ref{tab:envelope params}), we computed $M_{\mathrm{z}}$ at the abundance of their upper limits. The evolution of $M_{\mathrm{z}}$ for each metal is shown by the coloured solid lines in Figure~\ref{fig:parent body mass} as a function of time and calcium diffusion times ($\tau_\mathrm{{Ca}}$) since accretion ceased.

$M_{\mathrm{z,tot}}$ is shown by the black dashed line in Figure~\ref{fig:parent body mass} for each white dwarf. Based on the current atmospheric abundances in both stars, we estimated WD\,J2147$-$4035 has $M_{\mathrm{z,tot}} \approx 2 \times 10^{17}$\,g and WD\,J1922$+$0233 has $M_{\mathrm{z,tot}} \approx 6 \times 10^{17}$\,g at the present time. The mass of the accreted parent body increases exponentially with time elapsed since accretion stopped so this imposes an upper limit on the amount of diffusion times we can go back for each white dwarf; the bigger total mass the parent body has, the less likely it becomes that it could have been accreted. Therefore, it is unlikely that Neptune or Jupiter mass debris accreted onto WD\,J2147$-$4035 or WD\,J1922$+$0233. Also, the elemental abundances found in the atmospheres of these two stars (Table~\ref{tab:abundance_ratios}) are inconsistent with those of gas planets. We placed upper limits on $M_{\mathrm{z,tot}}$, and therefore $\tau_\mathrm{{Ca}}$, on WD\,J2147$-$4035 of approximately Earth mass ($\tau_\mathrm{{Ca}} \approx 17$) and WD\,J1922$+$0233 of approximately continental crust mass ($\tau_\mathrm{{Ca}} \approx 12$), due to the atmospheric chemical abundance ratios found in this star being most consistent with the Earth crust when considering the cessation of accretion (see Section~\ref{sec:atmos_abundance_ratios}).

We also determined the mass of hydrogen in the CVZ ($M_{\mathrm{H}}$) for both DZ stars, however the fraction of hydrogen mass accreted from planetary debris compared to that of a primordial origin from $M_{\mathrm{H}}$ cannot be unambiguously assigned. Thus, the mass of accreted hydrogen is not considered in $M_{\mathrm{z,tot}}$. Instead, $M_{\mathrm{H}}$ is reported for both DZ stars in Table~\ref{tab:envelope params}, in addition to $M_{\mathrm{z,tot}}$ and $M_{\mathrm{cvz}}/M_{\mathrm{WD}}$.

\subsubsection{Atmospheric chemical abundance ratios}
\label{sec:atmos_abundance_ratios}

\begin{figure*}
    \centering
    \subfigure[]{\includegraphics[width=\columnwidth]{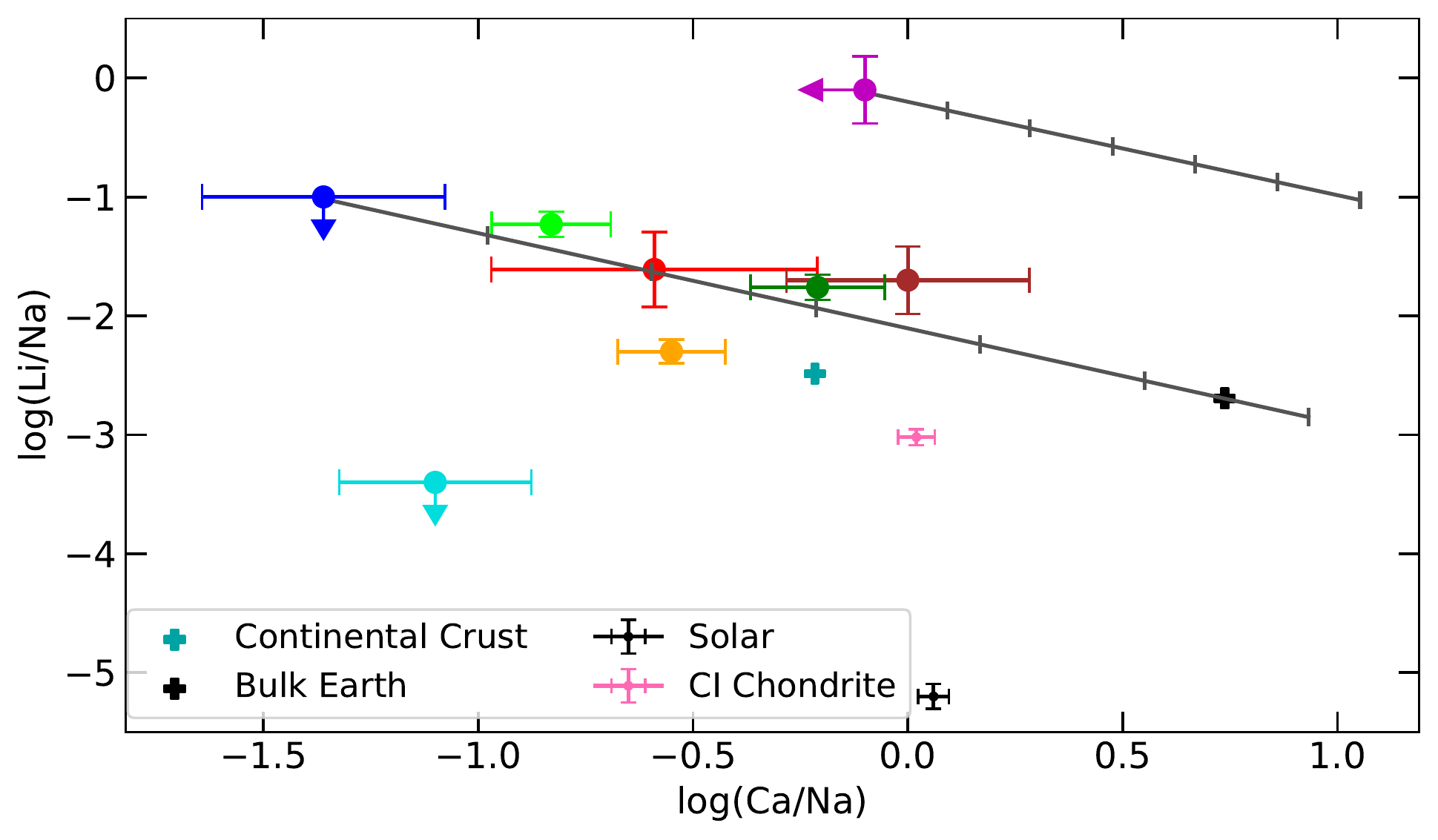}\label{fig:log(Ca/Na)}}
    \subfigure[]{\includegraphics[width=\columnwidth]{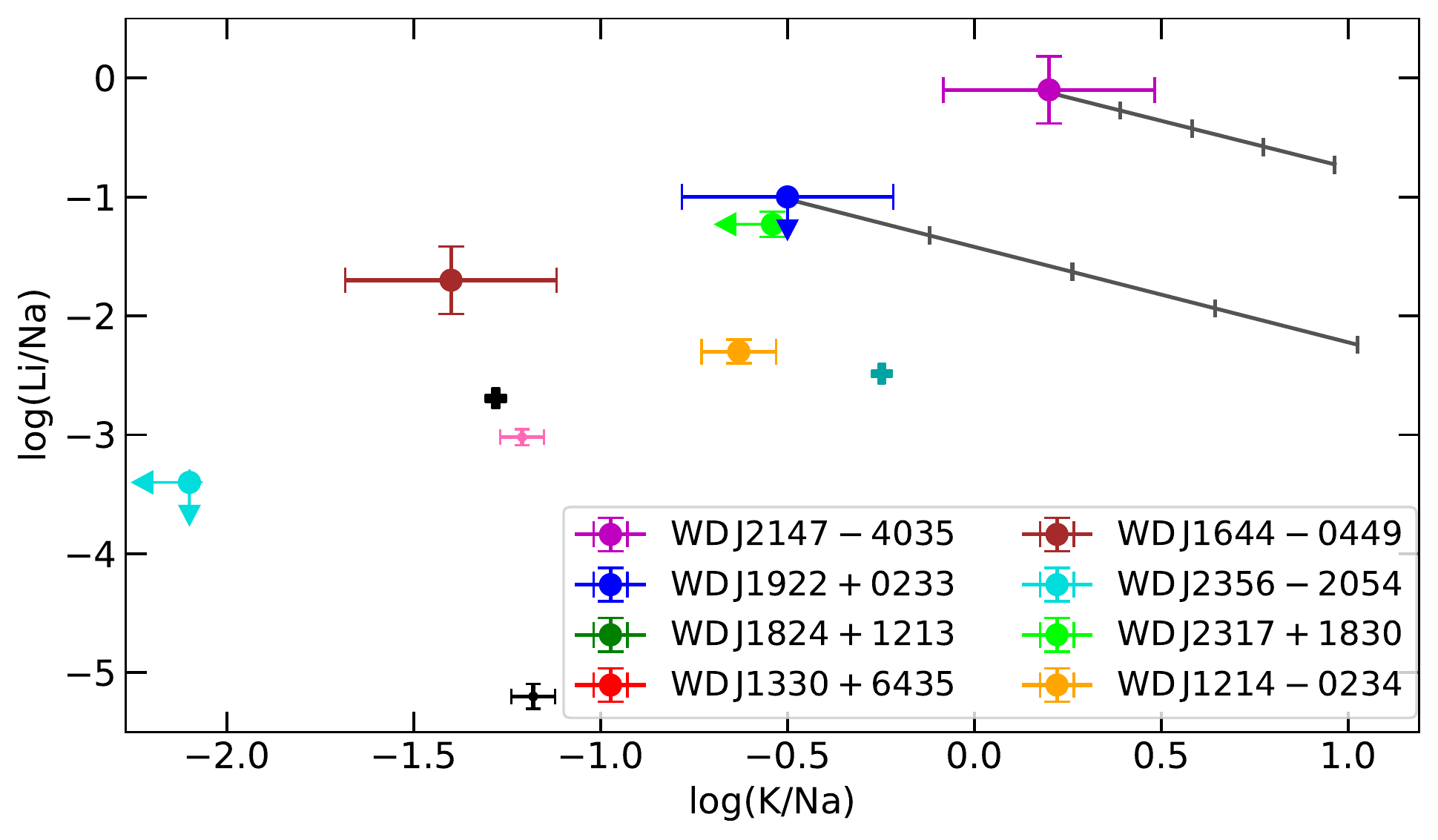}\label{fig:log(K/Na)}}
\caption{Abundance ratios for the DZ white dwarfs in our subsample compared to four Solar System benchmarks for (a) $\log\textnormal{(Li/Na)}$ versus $\log\textnormal{(Ca/Na)}$ and (b) $\log\textnormal{(Li/Na)}$ versus $\log\textnormal{(K/Na)}$. Error bars represent 1$\sigma$ uncertainties and are shown for all subsample objects except when only observational upper limits are derived, which are indicated by arrows. Diffusion timescales for WD\,J2147$-$4035 and WD\,J1922$+$0233 are shown by grey solid lines which indicate past photospheric abundances. Vertical grey markers on the diffusion timescales are in intervals of 1\,Myr for WD\,J2147$-$4035 and 5\,Myr for WD\,J1922$+$0233. The diffusion timescales for the previously published white dwarfs by \citet{Kaiser2021} and \citet{Hollands2021} are omitted but are approximately parallel to the tracks of WD\,J2147$-$4035 and  WD\,J1922$+$0233. The legends apply to both panels.}
\label{fig:abundance_plots}
\end{figure*}

To determine the elemental composition of the planetary debris which accreted onto WD\,J2147$-$4035 and WD\,J1922$+$0233, we calculated the current atmospheric abundance ratios (or observational upper limits) of $\log\textnormal{(Li/Na)}$, $\log\textnormal{(Ca/Na)}$ and $\log\textnormal{(K/Na)}$ (see Section \ref{sec:Spectroscopic_observations}) and compared them with Solar System benchmarks of bulk Earth \citep{mcdonough2000}, the continental crust \citep{Rudnick2003}, CI chondrites \citep{Lodders2003} and solar abundances \citep{Lodders2003} in Figure~\ref{fig:abundance_plots}. The other DZ stars in our subsample are included for context in Figure~\ref{fig:abundance_plots} with their published values shown in Table~\ref{tab:abundance_ratios}.

Our envelope code computed $\tau_{\mathrm{z}}$ (see Section~\ref{sec:accreted debris}) for each metal in the CVZ of WD\,J2147$-$4035 and WD\,J1922$+$0233 (Table~\ref{tab:envelope params}). Using the method outlined in \citet{Hollands2018_2}, we calculated the abundance histories for $\log\textnormal{(Li/Na)}$, $\log\textnormal{(Ca/Na)}$ and $\log\textnormal{(K/Na)}$ of both stars to analyse their past abundance evolutions. These evolutionary tracks are indicated by diagonal grey lines in Figure~\ref{fig:abundance_plots} and are marked in intervals of 1\,Myr for WD\,J2147$-$4035 and 5\,Myr for WD\,J1922$+$0233, due to its much slower diffusion times. Since the relative metal abundances currently observed in the atmospheres of WD\,J2147$-$4035 and WD\,J1922$+$0233 are exotic in comparison to Solar System benchmarks (see Figure~\ref{fig:abundance_plots}), and the fact it is not possible to know how long ago accretion ceased, we followed the evolutionary tracks of both stars to determine if their past atmospheric chemical abundance ratios represented parent bodies similar to the abundances of Solar System benchmarks after some elapsed diffusion times, while precluding to exceed the upper limits on $M_{\mathrm{z,tot}}$ we imposed on each star in Section~\ref{sec:accreted debris}. The abundance histories are not plotted for the other five published objects in our subsample \citep{Hollands2021, Kaiser2021} but are comparable to the evolutionary tracks for WD\,J2147$-$4035 and WD\,J1922$+$0233. 

Sodium produces the strongest absorption line in WD\,J2147$-$4035 and WD\,J1922$+$0233, yet these objects have the lowest $\log\textnormal{(Na/He)}$ abundances out of our DZ subsample (Table~\ref{tab:abundance_ratios}). The $\log\textnormal{(Ca/Na)}$ abundances for all white dwarfs in the subsample are similar or depleted compared to the Solar System benchmarks, with calcium not even being detected in WD\,J2147$-$4035. Solar System asteroids tend to have calcium to sodium ratios similar to unity or be slightly calcium-rich \citep[e.g.][]{Schramm1989, Zuckerman2007}. The atmospheric diffusion of calcium out of the CVZ is more rapid than sodium, therefore to explain these abundances the accretion events must have occurred a long time ago for a large fraction of calcium to have diffused towards the white dwarf core and be currently undetectable. For WD\,J1922$+$0233 to have a $\log\textnormal{(Ca/Na)}$ abundance comparable to the crust, accretion must have ceased some diffusion times ago due to its significantly depleted current ratio.

Lithium is not detected in WD\,J1922$+$0233, though we derived a $\log\textnormal{(Li/Na)}$ observational upper limit higher than Solar System benchmarks; however, its true value could be several 0.1\,dex lower which would make the $\log\textnormal{(Li/Na)}$ and $\log\textnormal{(Ca/Na)}$ abundances broadly consistent with the continental crust \citep{Rudnick2003} if accretion ceased $\approx 15$\,Myr ago. Providing the true $\log\textnormal{(Li/Na)}$ ratio inside the star is similar to our derived upper limit, WD\,J1922$+$0233 has an extremely similar composition to bulk Earth if accretion stopped $\approx 28$\,Myr ago (Figure~\ref{fig:log(Ca/Na)}). If debris with bulk Earth composition polluted WD\,J1922$+$0233 then its $M_{\mathrm{z,tot}}$ would have a comparable mass to the dwarf planet Ceres \citep[Figure~\ref{fig:parent body mass WDJ1922};][]{mccord2005ceres}. On the other hand, if a parent body with crust composition accreted $\approx 15$\,Myr ago, it would have $M_{\mathrm{z,tot}}$ comparable to Saturn's natural satellite Janus \citep{Jacobson2008}, which is $\approx 3$ orders of magnitude less massive than Ceres (Figure~\ref{fig:parent body mass WDJ1922}). The latter scenario is more likely as the recent accretion of smaller parent bodies is more probable than those of more massive bodies further in the past \citep{Hollands2021}. The accretion of a very small fraction of crust from an Earth-like planet is also probable as Earth's continental crust has a mass $\approx 4 \times 10^{25}$\,g \citep{Rudnick2003}. WD\,J1922$+$0233 has a much higher $\log\textnormal{(K/Na)}$ abundance than bulk Earth which only enhances with diffusion timescales, yet approaches the region of continental crust abundance $\approx 4$\,Myr ago, especially if the true lithium abundance is lower than the upper limit we derived. This decreases $M_{\mathrm{z,tot}}$ to $\approx 3$ orders of magnitude less massive than Janus and corresponds to an extremely small fraction of crust from an Earth-like planet. From the detected metals and corresponding past abundances in WD\,J1922$+$0233, we conclude it is probable that planetary bodies with crust-like compositions and $M_{\mathrm{z,tot}}$ consistent with, or $\approx 3$ orders of magnitude less than, Janus accreted $\approx 4-15$\,Myr ($\approx 1-7\,\tau_{\mathrm{Ca}}$) ago onto WD\,J1922$+$0233. This conclusion motivated us to set the undetected metals in our model atmospheres to Earth crust \citep{Rudnick2003} abundances relative to sodium (see Section~\ref{sec:Spectroscopic_observations}). We note that this star has a $\log\textnormal{(Al/He)}$ observational upper limit 0.64\,dex lower than the abundance found in the continental crust however this ratio increases to the continental crust value with elapsed diffusion timescales when considering accretion ceased in the past. Our atmospheric abundance assessment of WD\,J1922$+$0233 is based on current observations, yet additional follow-up spectroscopy could allow the calculation of more accurate observational upper limits and hence lead to a more conclusive debris analysis.

We also calculated the diffusion timescales of WD\,J1922$+$0233 using the $\Teff$, $\log g$, and \logHHe\ parameters computed by \citet{Bergeron2022} (see Section~\ref{sec:stellar_params}) to quantify the impact of this change on our accreted planetary debris conclusion. The interpretation of the abundance histories remains very similar, i.e. that planetary debris with a composition similar to the continental crust likely accreted onto WD\,J1922$+$0233 in the past. However, $\tau_{\mathrm{z}}$ of each metal significantly decreased when using the \citet{Bergeron2022} atmospheric parameters, resulting in a much more recent accretion event by $\approx 1-2$ orders of magnitude.

Calcium is not detected in WD\,J2147$-$4035, yet the observational upper limit of $\log\textnormal{(Ca/Na)}$ we derived is slightly higher than continental crust \citep{Rudnick2003} and could reduce to very similar values to the crust considering the true calcium abundance could be lower by several 0.1\,dex. The current $\log\textnormal{(Li/Na)}$ and $\log\textnormal{(K/Na)}$ abundances in WD\,J2147$-$4035 are significantly larger than the ratios found in Solar System benchmarks but are closest to continental crust composition. Accretion needed to have ceased $\approx 15$\,Myr ago to deplete the lithium abundance in WD\,J2147$-$4035 to the approximate continental crust $\log\textnormal{(Li/Na)}$ value, which would require a $M_{\mathrm{z,tot}}$ consistent with Earth's continental crust (Figure~\ref{fig:parent body mass WDJ2147}). However, this would enhance $\log\textnormal{(K/Na)}$ to 3.31\,dex higher than the ratio found in the crust. We also consider pollution via planetary debris enriched by primordial lithium, as in the \citet{Kaiser2021} scenario, in Section~\ref{sec:enhanced Li WDJ2147} to explain the high lithium abundance in WD\,J2147$-$4035, but we are not convinced this explains the high potassium abundance. The origin of carbon in this DZQH star is difficult to distinguish between accretion from external debris and convective dredge-up of core-carbon within the white dwarf, thus we omit carbon from our debris abundance analysis. We conclude the nature of planetary debris accreted by WD\,J2147$-$4035 remains elusive and more observations may be needed, including better limits on $\log\textnormal{(Ca/Na)}$, to provide more clarity on how long ago the accretion event occurred. However, we do not discount the possibility of multiple accretion events with parent bodies of varying calcium, lithium and potassium abundances at different times in its history.

\subsubsection{Primordial lithium enhancement in \texorpdfstring{WDJ\,2147$-$4035}{WDJ2147-4035}?}
\label{sec:enhanced Li WDJ2147}
High abundances of lithium are predicted to have been produced during Big Bang Nucleosynthesis (BBN), which occurred $\approx 3~-~30$\,minutes after the Big Bang \citep{Hou2017}. According to the \citet{Kaiser2021} scenario, the oldest stars and exoplanets would therefore be enriched by primordial lithium, which is then evident in the planetary debris observed in the atmospheres of the oldest white dwarfs. 

WD\,J2147$-$4035 has a large total age of $10.7\pm0.3$\,Gyr and we measured an extremely enhanced current atmospheric lithium abundance in this star, with a $\log\textnormal{(Li/Na)}$ ratio 2.38\,dex higher than the continental crust. However, WD\,J1824$+$1213 is a Galactic halo candidate with a likely large total age yet has a $\log\textnormal{(Li/Na)}$ ratio 1.66\,dex lower than WD\,J2147$-$4035. It is also unclear how this scenario could explain the extremely large $\log\textnormal{(K/Na)}$ ratio observed in WD\,J2147$-$4035 as potassium is not predicted to have been produced during BBN, but instead created in the cores of stars through stellar nucleosynthesis and dispersed throughout the Universe by supernovae \citep{Audouze1995, Tominaga2007, Iliadis2016}.

Observational detection limits and telluric lines from O$_2$ absorption by the Earth's atmosphere causes challenges and false-positives in the potassium detection of metal-poor stars with stellar spectroscopy \citep{Takeda2002, Takeda2009, Abohalima_Frebel2018}. Stellar model predictions of $\log\textnormal{(K/Fe)}$ in metal-poor stars thus tend to under-predict this ratio compared to stellar observations \citep[e.g.][]{Takeda2009}, though studies have found relations between metallicity ($\log\textnormal{(Fe/H)}$) and potassium abundance relative to the solar abundance \citep{Gratton1987b, Gratton1987, Chen2000, Takeda2002, Cayrel2004, Beers2005, Tominaga2007, Takeda2009}: an enhancement between $0 \lesssim \log\textnormal{(K/Fe)} \lesssim 0.3$\,dex from $\log\textnormal{(Fe/H)} \approx 0$\,dex to $\log\textnormal{(Fe/H)} \approx -1$\,dex for Galactic disc stars; then an approximately constant $\log\textnormal{(K/Fe)}$ ratio at lower metallicities between $-2.5 \lesssim \log\textnormal{(Fe/H)} \lesssim -1$\,dex consistent with halo stars; and a slight decrease in $\log\textnormal{(K/Fe)}$ for the oldest, hence extremely metal-poor (EMP; $-4 \lesssim \log\textnormal{(Fe/H)} \lesssim -2.5$\,dex) stars. \citet{Tominaga2007} report similar trends in the $\log\textnormal{(Na/Fe)}$ abundance relative to the solar abundance, but with an approximately constant abundance of $\log\textnormal{(Na/Fe)} \approx 0$\,dex for metallicities consistent with disc stars, a slight enhancement of $\log\textnormal{(Na/Fe)} \approx 0.5$\,dex for metallicities consistent with halo stars, then a decrease to $\log\textnormal{(Na/Fe)} \approx -0.8$\,dex in EMP stars. 

Therefore, the largest enhancement of $\log\textnormal{(K/Na)}$ compared to the solar abundance is observed in EMP stars, although there are uncertainties in these studies. The enhancement of $\log\textnormal{(K/Na)}$ for EMP stars is still not high enough to explain the extreme $\log\textnormal{(K/Na)}$ ratio currently seen in WD\,J2147$-$4035, plus the kinematics of WD\,J2147$-$4035 are inconsistent with EMP stars. The only other stellar population with a $\log\textnormal{(K/Na)}$ abundance enhancement relative to the solar abundance is old disc stars ($\log\textnormal{(Fe/H)} \approx -1$\,dex), although this enhancement is $\approx 1.1$\,dex too small to explain the abundance currently observed in WD\,J2147$-$4035.

\section{Conclusions}
\label{sec:Conclusions}
We have presented new spectroscopic observations of the ultra-cool DZ white dwarfs WD\,J2147$-$4035 and WD\,J1922$+$0233. These two stars occupy unusual positions on HRDs compared to objects in the \textit{Gaia} EDR3 white dwarf sample within 100\,pc of the Sun and the SDSS footprint \citep{GF2021} and the cool DZ subsample. WD\,J2147$-$4035 presents very red photometry as it has a depleted atmospheric hydrogen content compared to WD\,J1922$+$0233 and therefore has much milder CIA. Conversely, WD\,J1922$+$0233 exhibits unusually blue colours relative to its magnitude in Pan-STARRS and \textit{Gaia} EDR3 photometry for an ultra-cool star, due to strong atmospheric CIA causing the suppression of flux in the red optical and IR.

Our model atmosphere code used to fit WD\,J2147$-$4035 and WD\,J1922$+$0233 includes microphysics improvements in the non-ideal effects and treatment of CIA opacities, in addition to incorporating a reduction of the neutral line broadening of visible metals by an empirical factor of 100, compared to that of \citet{Koester2010}. This produced models that gave reasonable solutions for the observed spectroscopy and photometry of WD\,J2147$-$4035 and WD\,J1922$+$0233. Additional work still needs to be done to address uncertainties in the model atmospheres, such as understanding the behaviour of atoms and molecules at extreme atmospheric densities of $\approx 3\,\mathrm{g}\,\mathrm{cm}^{-3}$ to improve understanding of neutral line broadening and He-He-He CIA opacity, which would therefore improve future model atmospheres of ultra-cool DZ white dwarfs.

We found $\Teff = 3048\pm35$\,K for WD\,J2147$-$4035 and $\Teff~=~3343\pm54$\,K for WD\,J1922$+$0233, revealing them as the coolest and second coolest DZ white dwarfs known to date, respectively. WD\,J2147$-$4035 is also the intrinsically faintest confirmed white dwarf in the optical within the 40\,pc \textit{Gaia} sample (O'Brien et al., in preparation). The best-fitting \logHHe\ abundance ratio of WD\,J1922$+$0233 is close to the maximum intensity of H$_2$-He CIA which is consistent with observations of strong atmospheric CIA opacity in this white dwarf. The cooling age of WD\,J1922$+$0233 is $9.05\pm0.22$\,Gyr however its total age remains unknown because our model likely underestimated its mass, resulting in an unrealistically large main-sequence lifetime. WD\,J2147$-$4035 has a cooling age of $10.21\pm0.22$\,Gyr which is the largest known for a metal-polluted white dwarf. As this star does not show evidence of strong CIA, we were able to derive a total age for WD\,J2147$-$4035 of $10.7\pm0.3$\,Gyr. The kinematics of WD\,J2147$-$4035 and WD\,J1922$+$0233 suggest they are both Galactic disc candidates.

Strong sodium absorption lines are detected in WD\,J2147$-$4035 and WD\,J1922$+$0233, similar to the other white dwarfs in our DZ subsample. We also have firm detections of calcium in WD\,J1922$+$0233, lithium in WD\,J2147$-$4035 and potassium in both objects. We found WD\,J2147$-$4035 is magnetic due to the observed lithium spectral line being Zeeman split into three components and found a best-fitting magnetic field strength of $0.55\pm0.03$\,MG. Furthermore, we detected a photometric period of $\simeq13$\,h in the TESS FFI light curves of this star. Carbon is tentatively detected in WD\,J2147$-$4035 due to an excellent fit of the three strongest C$_2$ Swan band systems when distorted to the centroid wavelengths measured in the DQpecP star LP\,351$-$42, but further observations are needed to constrain the atmospheric carbon abundance. We assigned the spectral type DZQH to WD\,J2147$-$4035.

The current $\log\textnormal{(Ca/Na)}$ abundance in WD\,J1922$+$0233 is extremely low compared to Solar System benchmarks of continental crust, bulk Earth, solar and CI chondrites. However, considering the possibility that the accretion of debris onto this star has ceased, the past $\log\textnormal{(Ca/Na)}$ and $\log\textnormal{(Li/Na)}$ compositions approach similar values to Earth's continental crust $\approx 15$\,Myr ago -- especially if the true abundance of $\log\textnormal{(Li/Na)}$ is lower than the relatively high observational upper limit we derived. This is also true for the $\log\textnormal{(Li/Na)}$ and $\log\textnormal{(K/Na)}$ abundances in WD\,J1922$+$0233 from the accretion of a planetary body $\approx 4$\,Myr ago. The cessation of accretion $\approx 4-15$\,Myr ago corresponds to an accreted minimum parent body mass consistent with, or $\approx 3$ orders of magnitude less than, Janus, or an extremely small mass fraction of continental crust from an Earth-like planet. The detected metals and minimum parent body mass suggest this star was likely polluted by the accretion of planetary debris with a crust-like composition $\approx 4-15$\,Myr ago.

We find WD\,J2147$-$4035 has extremely enhanced $\log\textnormal{(K/Na)}$ and $\log\textnormal{(Li/Na)}$ abundances compared to Solar System benchmarks. Tracing the abundance evolution history of this star reveals many diffusion times are required to deplete the abundance of $\log\textnormal{(Li/Na)}$ to broadly approach the continental crust ratio, however $\log\textnormal{(K/Na)}$ would consequently increase to currently inexplicable values. Follow-up spectroscopy of WD\,J2147$-$4035 is required to further constrain the accretion history of this old, ultra-cool, magnetic, metal- and carbon-polluted white dwarf.

\section*{Acknowledgements}

This research received funding from the European
Research Council under the European Union’s Horizon 2020 research and
innovation programme number 101002408 (MOS100PC), the Leverhulme Trust Grant (ID RPG-2020-366) and the UK STFC consolidated grant ST/T000406/1.
Based on observations collected at the European Southern Observatory under ESO programme 105.20ET.001.
This work has made use of data from the European Space Agency (ESA) mission
{\it Gaia} (\url{https://www.cosmos.esa.int/gaia}), processed by the {\it Gaia}
Data Processing and Analysis Consortium (DPAC,
\url{https://www.cosmos.esa.int/web/gaia/dpac/consortium}). Funding for the DPAC
has been provided by national institutions, in particular the institutions
participating in the {\it Gaia} Multilateral Agreement.
The Pan-STARRS1 Surveys (PS1) and the PS1 public science archive have been made possible through contributions by the Institute for Astronomy, the University of Hawaii, the Pan-STARRS Project Office, the Max-Planck Society and its participating institutes, the Max Planck Institute for Astronomy, Heidelberg and the Max Planck Institute for Extraterrestrial Physics, Garching, The Johns Hopkins University, Durham University, the University of Edinburgh, the Queen's University Belfast, the Harvard-Smithsonian Center for Astrophysics, the Las Cumbres Observatory Global Telescope Network Incorporated, the National Central University of Taiwan, the Space Telescope Science Institute, the National Aeronautics and Space Administration under Grant No. NNX08AR22G issued through the Planetary Science Division of the NASA Science Mission Directorate, the National Science Foundation Grant No. AST-1238877, the University of Maryland, Eotvos Lorand University (ELTE), the Los Alamos National Laboratory, and the Gordon and Betty Moore Foundation.
This paper includes data collected by the TESS mission. Funding for the TESS mission is provided by the NASA's Science Mission Directorate.

\section*{Data Availability}
All data underlying this paper are publicly available from the relevant survey archives. The model atmospheres used in this paper are available upon reasonable request of the author, however the Koester model atmosphere codes and envelope codes are not publicly available.



\bibliographystyle{mnras}
\bibliography{ultracool_paper} 







\bsp	
\label{lastpage}
\end{document}